\documentclass[useAMS,usenatbib]{mn2e}

\usepackage[T1]{fontenc}
 \usepackage{pslatex}
\usepackage{amsmath}
\usepackage{amsfonts}
\usepackage{graphicx}
\usepackage{url}
\def\Blob{\mathop{\hbox{Blob}}}
\def\Exp{{\mathbb{E}}}
\usepackage{amsmath}
\usepackage{amssymb}
\usepackage{graphicx}
\def\citeapos#1{\citeauthor{#1}'s (\citeyear{#1})}
{\newif\ifnotend
\notendtrue
\def\veclist{ABCDEFGHIJKLMNOPQRSTUVWXYZabcdefghijklmnopqrstuvwxyz.}
\def\top#1#2.{#1}
\def\tail#1#2.{#2.}
\loop\expandafter\xdef\csname v\expandafter\top\veclist\endcsname%
{{\noexpand\bf\expandafter\top\veclist}}
\edef\veclist{\expandafter\tail\veclist}
\if\veclist.\notendfalse\fi\ifnotend\repeat}
\let\boldgrk=\gkvecten
\let\boldgrksc=\gkvecseven
\def\gkthing#1{{\mathchoice%
        {\hbox{{\boldgrk\char#1}}}
        {\hbox{{\boldgrk\char#1}}}
        {\hbox{{\boldgrksc\char#1}}}
        {\hbox{{\boldgrksc\char#1}}}}}

\def\valpha{\gkthing{11}}

\def\vtheta{\gkthing{18}}

\def\vpi{\gkthing{25}}

\def\vLambda{\gkthing{3}}

\def\d{{\rm d}}
\def\e{{\rm e}}
\def\pa{\partial}

\def\Dir{{\cal D}}
\def\pr{\mathop{\hbox{pr}}}
\def\tr{\mathop{\hbox{tr}}}
\def\diag{\mathop{\hbox{diag}}}

\title[Bayes vs.\ virial theorem]{Bayes versus the virial theorem:
  inferring the potential of a galaxy from a kinematical snapshot}
\author[J.~Magorrian]{John Magorrian\\
Rudolf Peierls Centre for Theoretical Physics, 1 Keble Road, Oxford
OX1 3NP}
\begin{document}

\date{}

\maketitle

\label{firstpage}

\begin{abstract}
  We present a new framework for estimating a galaxy's gravitational
  potential, $\Phi$, from its stellar kinematics
  by adopting a fully non-parametric model for the galaxy's unknown
  action-space distribution function, $f(\vJ)$.
  Having an expression for the joint likelihood of $\Phi$ and $f$, the
  likelihood of $\Phi$ is calculated by using a Dirichlet process
  mixture to represent the prior on $f$ and marginalising.
  We demonstrate that modelling machinery constructed using this
  framework is successful at recovering the potentials of some simple
  systems from perfect discrete kinematical data, a situation handled
  effortlessly by traditional moment-based methods, such as the virial
  theorem, but in which other, more modern, methods are less than
  satisfactory.  We show how to generalise the machinery to account
  for realistic observational errors and selection functions.  A
  practical implementation is likely to raise some interesting
  algorithmic and computational challenges.

\end{abstract}


\section{Introduction}

Inferring the mass distribution of a galaxy from limited information
on its stellar kinematics is a fundamental problem in modern
astrophysics.  Examples of this problem include estimating the masses
of central black holes and the properties of the dark matter haloes in
nearby galaxies from measurements of the integrated line-of-sight
velocity distributions
\citep[e.g.,][]{vdM+98,Siopis+09,Rix+97,Saglia+00,Thomas+07}.
Closer to home, surveys of the kinematical and chemical properties of
vast numbers of stars within our own Galaxy are becoming available
(see \cite{IvezicBeersJuric12} and \cite{RixBovy13} for recent
reviews), culminating in the Gaia mission \citep{Perryman+01} which
will provide positions and velocities for a sample of $\sim10^9$ stars.
A pressing challenge is to use such kinematical and chemical snapshots
to constrain the full dynamical structure of the Galaxy, including its
distribution of dark and luminous matter.

The problem addressed in this paper is the following: given some
stellar kinematical data $D$ and a list of gravitational potentials
$\Phi_1$, $\Phi_2$, ... corresponding to different assumed mass
distributions, how to calculate the likelihoods $\pr(D|\Phi_i)$?  For
simplicity, we may suppose that the galaxy under consideration is
collisionless and in a steady state with a single, chemically
homogeneous population of stars.  Then it is completely described by
just two unknown functions: its potential $\Phi(\vx)$ and the
distribution function $f(\vx,\vv)$ (hereafter DF) giving the
probability density of stars in phase space.
The problem becomes one of constraining $\Phi$ from observations that
probe only~$f$.  Jeans' theorem \citep{BT08} provides the crucial link
between these two unknown functions: in a steady-state galaxy,
$f(\vx,\vv)$ can depend on $(\vx,\vv)$ only through integrals of
motion in~$\Phi$.
It has long been known known that unwarranted assumptions about the
form of~$f$ can lead to incorrect conclusions about~$\Phi$
\citep[e.g.,][]{BinneyMamon82}.  Therefore any plausible scheme for
estimating~$\Phi$ must make minimal assumptions about~$f$.

There has been much previous work on this problem.
\cite{DejongheMerritt92} investigated the problem of constraining
$\Phi$ and~$f$ of a spherical galaxy given perfect knowledge of its
projected DF (i.e., its luminosity-weighted line-of-sight velocity
distribution).  They explained how $f$ could be reconstructed exactly
if $\Phi$ were known, and noted that the non-negativity constraint
$f\ge0$ allows many $\Phi$ to be ruled out.  A less idealised version
of the same problem was considered by \cite{MerrittSaha93}, who
developed an algorithm for assigning likelihoods to spherical
potentials given projected positions and radial velocities for a
discrete sample of stars.  Even less idealised variants of the same
problem come from investigating how well one can estimate the masses
of galaxies' central black holes \citep[e.g.,][]{VME04} or dark-matter
haloes \citep[e.g.,][]{Gerhard+98} from noisy, integrated kinematics
that have finite spatial and velocity resolution.  Apart from
\cite{DejongheMerritt92}, all of these methods identify a
single preferred $f=f_{\rm best}$ for each trial~$\Phi$ and assign
$\pr(D|\Phi)=\pr(D|\Phi,f_{\rm best})$, an assumption that has been
questioned by \cite{magog06}.

In the present paper I revert to an extremely idealised situation in
which the data $D$ represent an unbiased sample of the galaxy's stars,
for each of which we know $(\vx,\vv)$ precisely.  Then the problem
becomes one of inferring $\Phi$ given a random realisation of~$f$.
The most obvious way of tackling this is by applying moment-based
methods, such as the virial theorem.  Unfortunately, moment-based
methods are not easy to extend to the general case of imprecise
measurements with complicated selection effects.  My motivation for
the paper was to find a coherent alternative to the virial theorem and
its variants that can naturally be extended to allow the computation
of $\pr(D|\Phi)$ in more realistic scenarios.

Notice that -- even in this extremely idealised scenario -- we do not
know the DF directly, but instead have only a random realisation of
it.  This suggests that we treat~$f$ as a nuisance function that is to
be marginalised: the desired $\pr(D|\Phi)$ is then obtained by
integrating the well-defined joint likelihood $\pr(D|\Phi,f)$ over all
possible~$f$, the contribution from each~$f$ weighted by a prior that
must satisfy certain consistency conditions.  From the statistics and
machine-learning communities I borrow the idea of using a {\it
  Dirichlet process mixture} \citep[e.g.,][]{Teh10} to model the prior
distribution on the DF.  In effect, the DF is modelled as a
distribution of an arbitrary number of blobs of arbitrary size and
shape in action space with a suitably chosen prior for the
distribution of blob locations, shapes, sizes and weights.

The paper is organised as follows.  Section~\ref{sec:toy} uses a toy
one-dimensional problem to underscore some of the shortcomings of
existing methods for computing $\pr(D|\Phi)$, particularly for the
idealised case of perfect (or very good) data.
Section~\ref{sec:solution} sets out the core ideas of the proposed
solution.  It introduces the idea of a Dirichlet process mixture and
explains how, by treating the distribution of possible DFs as such a
mixture, one can calculate $\pr(D|\Phi)$ by marginalising the joint
likelihood $\pr(D|\Phi,f)$ over~$f$.  The technical details of two
different schemes for carrying out this marginalisation are relegated
to appendices.
Section~\ref{sec:applications} demonstrates that this idea works by
applying it to some simple test problems.  Its relation to some other
potential-estimation methods is discussed in
Section~\ref{sec:discuss}, while Section~\ref{sec:gener} explains how it
can be extended to take proper account of the observational errors and
selection biases in real catalogues.  Section~\ref{sec:summary} sums
up,

\section{A toy problem}
\label{sec:toy}

Consider a one dimensional galaxy in which stars move in a potential
$\Phi(x)=\frac12\omega^2x^2$ and have some unknown distribution
function $f(J)$, where the action $J=\omega(x^2+v^2/\omega^2)/2\pi$.
Given a sample, $D$, consisting of the positions and velocities
$(x^\star_n,v^\star_n)$ of $N$ stars drawn from this galaxy, what
constraints can we place on $\omega$?  In particular, what is the
posterior probability distribution $\pr(\omega|D)$?  We assume
that there are no selection effects -- the sample $D$ is a
fair representation of the underlying DF -- and that we know the
$N$ stars' positions and velocities precisely.

\subsection{Virial theorem}

The virial theorem provides an effortless solution to this problem.
The DF $f$ satisfies the collisionless Boltzmann equation,
\begin{equation}
  \label{eq:CBE}
  \frac{\pa f}{\pa t}+v\frac{\pa f}{\pa x}
  -\frac{\pa\Phi}{\pa x}\frac{\pa f}{\pa v}=0.
\end{equation}
Assume that the galaxy is in a steady state, so that $\pa f/\pa
t=0$, and that $f$ tapers off smoothly to zero for large $|x|$ and
$|v|$.  Multiplying~\eqref{eq:CBE} by $xv$, integrating over the
$(x,v)$ phase plane and rearranging gives
\begin{equation}
  \omega^2=\frac{\int f v^2\,\d x\d v}{\int f x^2\d x\d v},
\label{eq:VT1}
\end{equation}
which, as our $N$ stars provide a fair sample of $f$, can be
estimated as
\begin{equation}
  \omega^2\simeq\omega_{\rm VT}^2\equiv\frac{\sum_{n=1}^N{v_n^\star}^2}{\sum_{n=1}^N{x_n^\star}^2}.
\label{eq:VT2}
\end{equation}
This approach is straightforward and to the point, but it suffers from
the following drawbacks.
\begin{enumerate}

\item Going from~\eqref{eq:VT1} to~\eqref{eq:VT2} involves estimating
  integrals over the DF by taking appropriately weighted sums of the
  observed star distribution.  These estimates ignore the strong
  constraint on the DF provided by Jeans' theorem: when viewed as a
  function of action--angle coordinates $(J,\theta)$ instead of
  $(x,v)$, the DF $f=f(J)$ must be uniform in angle.  Therefore,
  although $\omega_{\rm VT}\to\omega$ in the limit $N\to\infty$, we
  should be able to do better for finite values of $N$.  As an extreme
  example, given a sample of, say, $N=4$ stars that all happen to lie
  exactly on an ellipse ${x_n^\star}^2+{v_n^\star}^2/\omega_0^2=1$ for some
  $\omega_0$, it is more plausible to believe $\omega=\omega_0$ over
  whatever estimate $\omega_{\rm VT}$ provides.

\item Apart from some special cases \citep[e.g.,][and references
  therein]{AnEvans11}, there is no general way of modifying the
  moment-based estimate~\eqref{eq:VT2} to take account of
  uncertainties in measurements of the phase-space
  coordinates~$(\vx^\star,\vv^\star)$.  For example, in the Milky Way
  one typically has only very crude estimates of the distances to
  individual stars, which in turn affects the estimate of their
  transverse velocities from their proper motions.
\item Real stellar catalogues rarely provide a fair, unbiased sample
  of the DF underlying the galaxy.  Observations are inevitably
  subject to some selection function $S(\vx,\vv)$, which gives the
  probability that a star at $(\vx,\vv)$ would be included in the
  sample.  Although it is possible to extend the analysis above to use
  an ``selective DF'' $f_{\rm S}(\vx,\vv)=S(\vx,\vv)f(\vx,\vv)$, the
  results are dominated by any sharp features in $S(\vx,\vv)$.  In
  other words, they are strongly affected by what is happening at the
  edges of the survey, which is worrying as one rarely knows
  $S(\vx,\vv)$ well.
\end{enumerate}

\subsection{The modern approach: maximum-likelihood orbit-based models}
\label{sec:osmodels}
The most flexible modern scheme for estimating the potential is the
so-called ``orbit superposition'' or ``extended-Schwarzschild'' method
and its variants (see \cite{Chaname+08} for an
application to discrete kinematics).  These work by considering a
range of explicitly chosen trial potentials $\Phi$.  For each such
$\Phi$ they:
\begin{enumerate}
\item Represent the DF as a weighted sum $f(\vJ)=\sum_kw_kf_k(\vJ)$ of
  basis functions $f_k(\vJ)$ that depend only on integrals of motion
  in the assumed $\Phi$.  This ensures that Jeans' theorem is
  satisfied.  The simplest way of constructing such a basis is to take
  a representative sample of single orbits in $\Phi$
  \citep{Schwarzschild79}.  The next simplest is to represent each $f_k$
  by a bunch of neighbouring orbits.
\item Calculate the contribution
  $P_{nk}=\pr((x^\star_n,v^\star_n)|f_k,\Phi)$ that each basis element
  makes to each observed datapoint, including the effects of any
  observational uncertainties.
\item Find the set of weights $w_i$ that maximises the likelihood
  $\pr(D|\{w_k\},\Phi)$.  For the present problem this likelihood is
  $\prod_n\sum_kP_{nk}w_k$ and is subject to the 
  constraint that $\sum_kw_k=1$.
\item Take this peak value of the likelihood as the likelihood
  $\pr(D|\Phi)$ of the trial potential $\Phi$.
\end{enumerate}

\begin{figure}
  \centering
  \includegraphics[width=0.8\hsize]{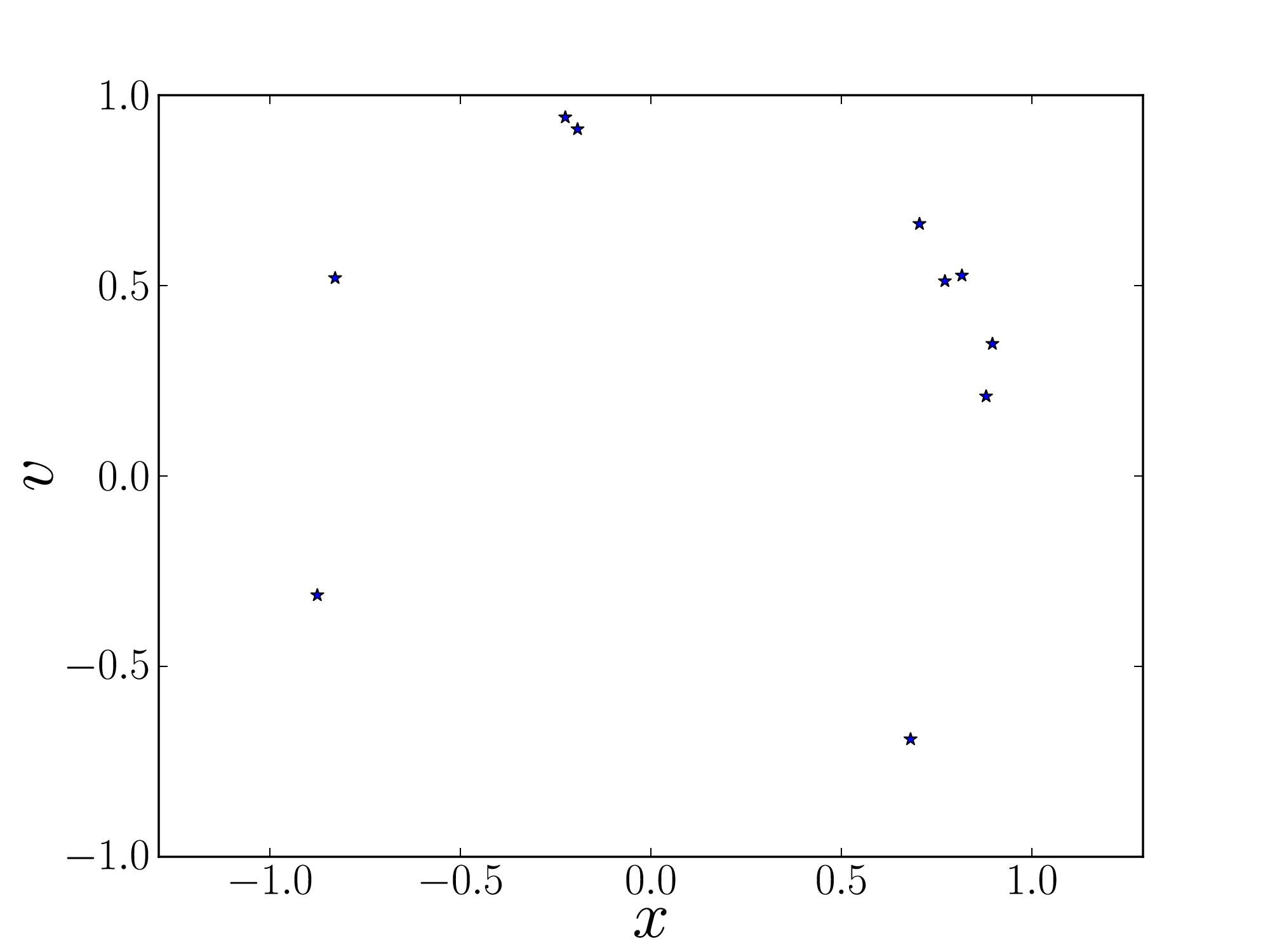}

  \caption{A sample of $N=10$ stars drawn from a one-dimensional
    toy galaxy with a simple harmonic oscillator potential
    $\Phi(x)=\frac12\omega_0^2x^2$ in which $\omega_0=1$.  The
    underlying DF of the model is a uniform distribution in amplitude
    $a$ between $a=0.9$ and $a=1$, where
    $a^2=x^2+v^2/\omega_0^2$.}
\label{fig:exampleD}
\end{figure}
\begin{figure}
  \centering
  \includegraphics[width=0.8\hsize]{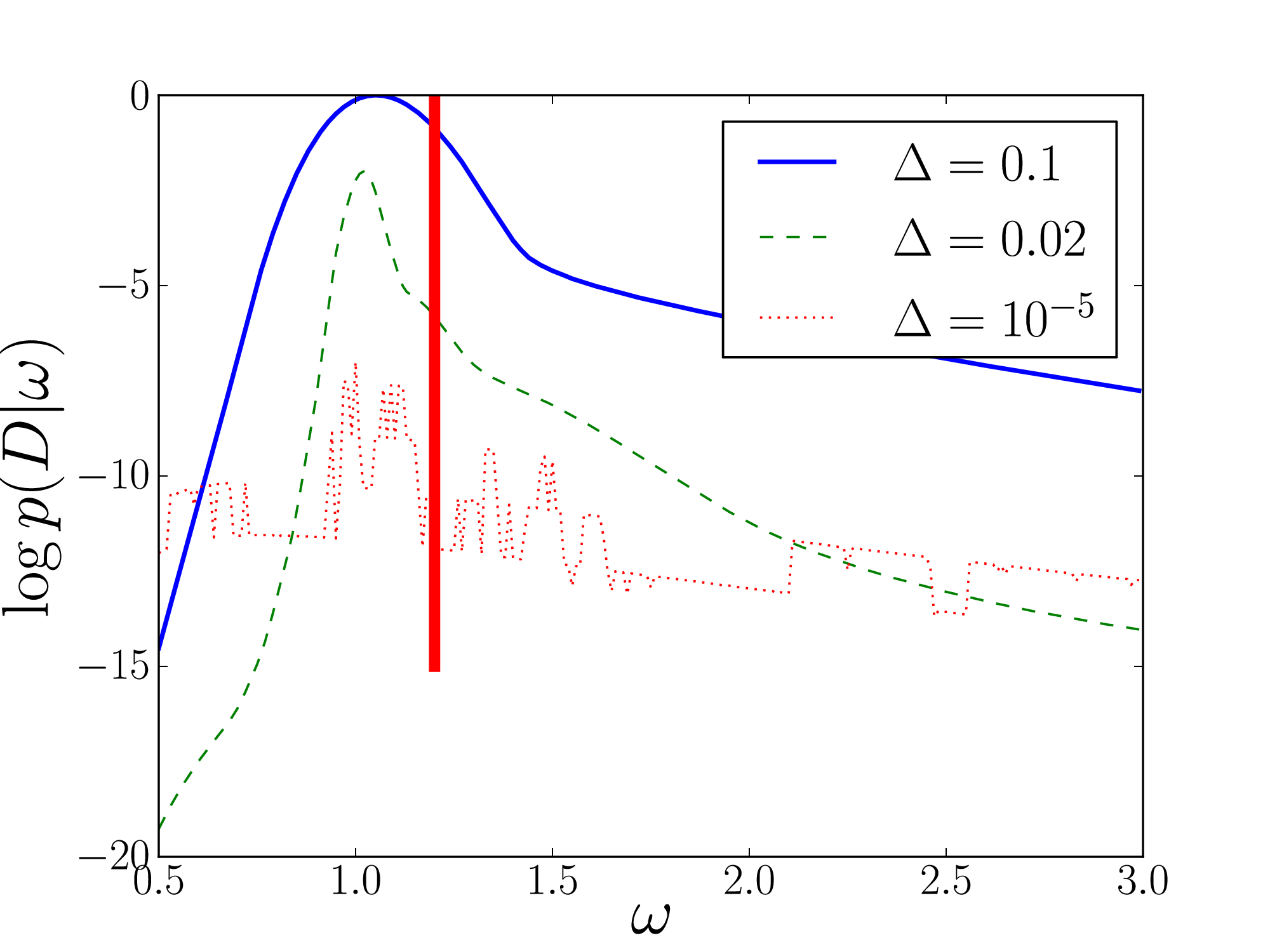}

  \caption{The likelihood $\pr(D|\omega)$ for the sample of 10 stars
    shown in Figure~\ref{fig:exampleD} calculated using the standard
    maximum likelihood-based algorithm by adding nominal Gaussian error circles
    of standard deviation $\Delta$ around each observed $(x,v)$.
    The results for different choices of $\Delta$ have been offset for
    clarity.  As
    $\Delta\to0$ the models cannot distinguish one potential from
    another.  For comparison, the heavy vertical red line indicates
    the estimate~\eqref{eq:VT2} of $\omega$ obtained by using the
    virial theorem.}
\label{fig:osresults}
\end{figure}
Let us see how well such a scheme works when applied to our toy
one-dimensional problem.  Figure~\ref{fig:exampleD} shows a sample of
$N=10$ stars drawn from a galaxy with $\omega_0=1$ and the annular top-hat DF
\begin{equation}
  f(J)=
  \begin{cases}
    A, & \hbox{if $0.9<a(x,v)<1$,}\\
    0, & \hbox{otherwise},
  \end{cases}
\end{equation}
where $A>0$ is an uninteresting normalisation constant and the
amplitude $a(x,v)$ of an orbit passing through $(x,v)$ is defined to be
\begin{equation}
  a^2(x,v)\equiv x^2+\frac{v^2}{\omega^2}=\frac{2\pi J}\omega,
\end{equation}
which for visualisation purposes is more convenient to use than the action
$J$ when labelling orbits on the phase plane.
To illustrate the effects of observational uncertainties, we assign
nominal Gaussian errors of standard deviation $\Delta$ to each
$(x^\star_n,v^\star_n)$ and consider the effects of shrinking $\Delta$
towards zero.

We use the four-step modelling procedure above to assign a likelihood
$\pr(D|\omega)$ to each of a range of trial values of $\omega$.  The
DF is modelled by a set of abutting annuli, with
$f_k(x,v)=\hbox{constant}$ for amplitudes
$a_k^2<x^2+v^2/\omega^2<a_{k+1}^2$, zero otherwise.  There are $200$
such annuli, running from $a_1=0$ to $a_{201}=2$ with uniform spacing
$a_{k+1}-a_k=0.01$.  The contribution $P_{nk}$ that the $k^{\rm th}$
such annulus makes to the probability of observing the $n^{\rm th}$
star is simply the integral of $f_k(x,v)$ times a Gaussian of width
$\Delta$ centred on $(x^\star_n,v^\star_n)$.  Having calculated these
$P_{nk}$, we use the expectation--maximisation algorithm to find the
set of weights $w_k$ that maximise the likelihood subject to the
constraint that $\sum_kw_k=1$

The resulting plot of maximum likelihood versus assumed $\omega$ for
this dataset is shown on Figure~\ref{fig:osresults}.  When $\Delta$ is
large, this procedure produces a likelihood distribution
$\pr(D|\omega)$ that peaks close to the correct value of $\omega_0=1$.
Perversely, however, the likelihood flattens as $\Delta$ shrinks: if
the data become {\it too} good, the model is unable to distinguish one
potential from another!  It is easy to see why this is: given perfect
knowledge of $(x^\star_n,v^\star_n)$ there is a unique orbit in
(almost) any given potential that passes through this
$(x^\star_n,v^\star_n)$ and no other; it is only when several stars
lie along an orbit that one can say anything about the likelihood of
the assumed potential.  Therefore all potentials have the same
likelihood.

\subsection{Comments}

What to do about this?  One remedy is to consider only strongly
parametrised forms for the DF or to take a non-parametric DF and
impose some form of regularisation \citep[e.g.,][]{Merritt93}, but this
has the disadvantage of introducing hard-to-understand biases in
$\pr(D|\Phi)$.  A related idea would be to somehow couple the resolution of
the basis functions $f_k$ to some properties of the available sample
$D$.
Fundamentally, however,
the problem with the maximum-likelihood procedure above is that it
looks only at the distribution of orbits that produces the best
possible match to the observed sample, with no regard for nearby orbit
distributions that are only slightly less likely.

\section{Modelling the distribution of DFs as a Dirichlet process mixture}
\label{sec:solution}

Here is a more general restatement of the toy problem above.  We have
a galaxy with unknown potential $\Phi(\vx)$ and unknown DF
$f(\vx,\vv)$.  We are given a list of the phase-space locations
$(\vx^\star_n,\vv^\star_n)$ of $N$ stars drawn from the galaxy.  We may
assume that the galaxy is in a steady state and that the list of stars
is a fair sample of the underlying DF.  Our job is to constrain the
potential from these data~$D$.  In particular, we seek the posterior
probability distribution $\pr(\Phi|D)$.  By Bayes' theorem
$\pr(\Phi|D)$ is proportional to $\pr(D|\Phi)\pr(\Phi)$, where
$\pr(\Phi)$ is our prior on $\Phi$.

As the galaxy is in a steady state, it is natural to express the DF in
terms of action--angle coordinates $(\vJ,\vtheta)$ instead of
$(\vx,\vv)$: by the strong Jeans theorem, the DF is a function
$f(\vJ)$ of the actions only (BT08).  Let $d=1$, 2 or 3 be the number
of dimensions in the system.  If $d=3$ then we may take
$\vJ=(J_r,J_\theta,J_\phi)$ in which the radial action $J_r$ and the
latitudinal action $J_\theta$ must be non-negative.  The azimuthal
action $J_\phi$ can take either sign.  Similarly, for $d=2$ we have
$\vJ=(J_r,J_\phi)$ in which $J_r\ge0$, while for $d=1$ the single action $J\ge0$.
The likelihood $\pr(D|\Phi)$ can be
expressed in terms of the stars' actions $\vJ^\star_1...\vJ^\star_N$
as
\def\assumptions{{\cal A}}
\def\cellparameters{{\cal C}}
\begin{equation}
\label{eq:PDgivenPhi}
\begin{split}
\pr(D|\Phi,\assumptions)
&=\int\d^d\vJ^\star_1\d^d\vtheta^\star_1
   \pr(\vx^\star_1,\vv^\star_1|\vJ_1^\star,\vtheta^\star_1,\Phi)
\cdots\\
&\int\d^d\vJ^\star_N\d^d\vtheta^\star_N
  \pr(\vx^\star_N,\vv^\star_N|\vJ_N^\star,\vtheta^\star_N,\Phi)
\,\cdot\,
\pr(\vJ_1^\star\cdots\vJ_N^\star|\assumptions),
\end{split}
\end{equation}
where $\assumptions$ denotes some as-yet unstated assumptions
(which will be summarised in \S\ref{sec:DPMsummary} below) and each
$\pr(\vx^\star_n,\vv^\star_n|\vJ^\star_n,\vtheta^\star_n,\Phi)$ is
simply a Dirac delta that picks out the $\vJ^\star_n$
corresponding to $(\vx^\star_n,\vv^\star_n)$ for the assumed~$\Phi$.
The only place that the potential enters into this problem is in the
conversion from $(\vx^\star_n,\vv^\star_n)$ to $(\vJ^\star_n,\vtheta^\star_n)$;
when expressed in terms of the actions, the likelihood
$\pr(\vJ^\star_1...\vJ^\star_N|\assumptions)$ is independent
of~$\Phi$.

The DF does not appear explicitly in the innocuous-looking expression
$\pr(\vJ^\star_1...\vJ^\star_N|\assumptions)$ because it has been
marginalised out: we have that 
\begin{equation}
  \pr(\vJ^\star_1...\vJ^\star_N|\assumptions)
=\int\pr(\vJ^\star_1...\vJ^\star_N|f,\assumptions)\pr(\d f|\assumptions)
\label{eq:sumoverDF}
\end{equation}
which involves summing the likelihood
$\pr(\vJ^\star_1...\vJ^\star_N|f,\assumptions)$ over all DFs $f(\vJ)$
that satisfy the uniform-in-angle constraint imposed by Jeans'
theorem.  There remains the choice of prior measure $\pr(\d
f|\assumptions)$, which is a distribution over distributions.
A standard way of choosing this, well known from the statistics and
machine-learning communities, is to model the DF as being drawn from a
Dirichlet process mixture: essentially, $f(\vJ)$ is expressed as a sum
of an arbitrary number of blobs in action space, the blobs having some
distribution of locations, sizes, shapes and probability masses.
The marginal likelihood~\eqref{eq:sumoverDF} is obtained by
marginalising the parameters that describe the blobs.  This basic idea
is explained more precisely below, with further discussion postponed
until Section~\ref{sec:discuss}.

In the following let 
$V$ be a large, but finite, volume of action space $(J_{\rm box})^d$ that includes
all of the $\vJ_n^\star$ and let $H$ be a measure on this space.  We
take $H$ to be proportional to the canonical $(2\pi)^d\d^d\vJ$
phase-space volume, normalised so that $H(V)=1$.  

\subsection{Dirichlet distribution}

Consider an arbitrary partition~$P$ of action space~$V$ into an
arbitrary number~$K$ of cells.  Let $V_k$ be the subvolume enclosed by
the $k^{\rm th}$ cell and let $\pi_k$ be the associated probability
mass: that is, $\pi_k$ is an integral over the unknown DF
within~$V_k$.  As the DF is unknown, we may treat
$\vpi=(\pi_1,...,\pi_K)$ as a list of random variables.  Clearly the
$\pi_k$ must satisfy the conditions $\pi_k\ge0$ and
$\sum_{k=1}^K\pi_k=1$.  For simplicity, let us assume that the $\pi_k$
are independent of one another.  This is a strong assumption, whose
consequences are discussed at the end of section~\ref{sec:DP} and
further in section~\ref{sec:discuss} below.

Recognising that the choice of partition $P$ is arbitrary yields an
important constraint on the prior~$\pr(\vpi)$.  Given any $P$, we may
construct a new partition~$P'$ by merging, say, the first two cells of
$P$ together, so that the volume of the first cell of~$P'$ is
$V_{12}=V_1\cup V_2$, with associated probability mass
$\pi_{12}=\pi_1+\pi_2$.  Conversely, given $P'$ we can construct $P$
by splitting one of the cells of $P'$ into two.  For consistency, the
prior on $P'$ should be related to the prior on $P$ through
\begin{equation}
  \begin{split}
    \pr(\pi_{12},\pi_3,...,\pi_K|P')&=\int\d\pi_1\int\d\pi_2\,
    \delta(\pi_1+\pi_2-\pi_{12})\times\\
    &\quad\qquad\pr(\pi_1,\pi_2,\pi_3,...,\pi_K|P).\\
  \end{split}
\label{eq:agglom}
\end{equation}
A particularly simple form for the prior that satisfies these
conditions is the Dirichlet distribution, which has probability
density function
\begin{equation}
\Dir(\vpi|\valpha)
\equiv \delta\left(1-\sum_{l=1}^K\pi_l\right)
C(\valpha)
\prod_{k=1}^K\pi_k^{-1+\alpha_k},\label{eq:dirichlet}
\end{equation}
where the free parameters $\valpha=(\alpha_1,...,\alpha_K)$ satisfy
$\alpha_k>0$, and the normalising constant
\begin{equation}
  \label{eq:Cdefn}
  C(\valpha) \equiv \frac{\Gamma\left(\sum_{k=1}^K\alpha_k\right)}
  {\prod_{k=1}^K\Gamma(\alpha_k)},
\end{equation}
with $\Gamma(\alpha)$ the usual Gamma function.  A convenient
shorthand for~\eqref{eq:dirichlet} is
\begin{equation}
  (\pi_1,...,\pi_K)\sim \Dir(\alpha_1,...,\alpha_K),
\end{equation}
the $\sim$ sign here meaning ``is distributed as''.  
It is not hard to show that, if
\begin{equation}
  (\pi_1,\pi_2,\ldots,\pi_K)\sim\Dir(\alpha_1,\alpha_2,\alpha_3,\ldots,\alpha_K),
\end{equation}
then
\begin{equation}
  (\pi_{12},\pi_3,\ldots,\pi_K)\sim\Dir(\alpha_1+\alpha_2,\alpha_3,\ldots,\alpha_K).
\end{equation}
Therefore the prior~\eqref{eq:dirichlet} sastifies the consistency
condition~\eqref{eq:agglom} as long as we choose the coefficients
$\alpha_k$ proportional to the volume measure $H(V_k)$ associated with
each cell.

\subsection{Dirichlet process}
\label{sec:DP}
The consistency condition~\eqref{eq:agglom} means that we may restrict
our attention in the following to priors defined on a large number $K$
of very small cells that all have the same volume and differ only
their locations $\vJ_k$ in action space.  As the cells have identical
volumes, they must also have identical values of $\alpha_k$.  So, let
us take $\alpha_k=\alpha/K$ and consider the limit $K\to\infty$
\citep{Neal00,Rasmussen00}.

Any partition of $V$ into a finite number of nonempty, non-overlapping
subvolumes, $(V_1,...,V_L)$, can be represented by grouping together
these tiny, equal-volume cells; each of the $K$ cells will lie inside
precisely one of the $V_l$.  Let $F(V_l)$ be the probability mass
associated with $V_l$, so that 
$F(V_l)=\sum_{V_k\in V_l}\pi_k$.
Using the consistency property~\eqref{eq:agglom} of the Dirichlet
distribution~\eqref{eq:dirichlet} together with the choice
$\alpha(V_l)=\alpha H(V_l)$, it is obvious that 
for any such partition
$(V_1,...,V_L)$ of $V$ we have that
\begin{equation}
  \label{eq:DPdefn}
  (F(V_1),...,F(V_L))\sim \Dir(\alpha H(V_1),...,\alpha H(V_L)).
\end{equation}
This is the defining property of a {\it Dirichlet process}
(\cite{Ferguson73}; see also \cite{Teh10} for a brief, accessible
introduction).  A Dirichlet process has two parameters.  One is the
base measure $H$, which we take to be proportional to the canonical
volume element $(2\pi)^d\d^d\vJ$.  The other is the {\it concentration
  parameter}~$\alpha$, which controls the clumpiness of the
distribution: the expectation value of $F(V_l)$ is just $H(V_l)$ and
the variance is $H(V_l)(1-H(V_l))/(\alpha+1)$; as $\alpha$ increases
the variance shrinks.

To understand more about the properties of draws from a Dirichlet
process and the effect of~$\alpha$, let us return to the picture of
the limit of a large number $K\to\infty$ of equal-volume cells and
suppose we draw $N$ stars from the distribution~\eqref{eq:dirichlet}.
Let $c_i\in\{1,...,K\}$ be the cell number of the $i^{\rm th}$ star.
Clearly, $\pr(c_i=c)=\pi_{c}$: the probability that the $i^{\rm th}$
draw picks cell~$c$ is just $\pi_c$.  Marginalising
$\vpi=(\pi_1,...,\pi_K)$ with the prior~\eqref{eq:dirichlet}, the
probability of drawing star~1 from cell $c_1$, ..., star~$N$ from cell
$c_N$ is
\begin{equation}
  \begin{split}
    &\pr(c_1...c_N)\\&=
  \frac{\Gamma(\alpha)}{\left[\Gamma\left(\frac\alpha  K\right)\right]^K}
  \int\d\vpi\delta\left(1-\sum_{l=1}^K\pi_l\right)\,
  \pi_{c_1}\cdots\pi_{c_N}\pi_1^{-1+\alpha/K}\cdots\pi_K^{-1+\alpha/K}\\
  &=\frac{\Gamma(\alpha)}{\left[\Gamma\left(\frac\alpha
        K\right)\right]^K\Gamma(N+\alpha)}
  \prod_{k=1}^K{\Gamma\left(n_k(N)+\frac\alpha K\right)},
  \end{split}
\end{equation}
where $n_k(N)$ is the number of stars in cell $k$ for this draw of $N$
stars.  Therefore the conditional probability
\begin{equation}
  \pr(c_N=c|c_1,...,c_{N-1})=\frac{\pr(c_1...c_N)}{\pr(c_1...c_{N-1})}
    =\frac{n_k(N)+\frac\alpha K}{N-1+\alpha}.
\end{equation}
So, the first star is equally likely to come from any of the $K$
cells.  In the limit $K\to\infty$ the second star has probability
$1/(1+\alpha)$ of coming from the same cell as the first star.  The
remaining probability $\alpha/(1+\alpha)$ is spread equally among the
unoccupied cells.  More generally, star~$N$ has probability
$n/(N-1+\alpha)$ of coming from a cell that already holds $n$ stars.
The probability that it does not come from a cell occupied by any
of the previous $N-1$ stars is $\alpha/(N-1+\alpha)$.  The same
behaviour can be derived directly from the more abstract
definition~\eqref{eq:DPdefn}.  When $N$ is large the expectation value
of the number of non-empty cells tends to $\alpha\log (1+N/\alpha)$
\citep[e.g.,][]{Antoniak74,Teh10}.


This argument shows that if we use the $\pi_k$ to represent the DF
directly the DF will be a series of isolated spikes: neighbouring
parts of action space do not ``know'' about each other.  Clearly then
the marginal likelihood~\eqref{eq:sumoverDF} would be independent of
how the $\vJ_n^\star$ are distributed, unless two or more of them
happen to overlap precisely.  Therefore $\pr(D|\Phi)$ would be flat,
just as we found for the models in Figure~\ref{fig:osresults}.
In fact, {\it any} prior that treats the $\pi_k$ as independent
variables subject to the consistency condition~\eqref{eq:agglom} will
produce spiky, discrete DFs \citep{Kingman92} and therefore will suffer
from this problem.

\subsection{Dirichlet process mixture of blobs}

\label{sec:DPM}
In order to give the prior on the DF some notion of continuity, let us
smear out the probability mass $\pi_k$ associated with the $k^{\rm
  th}$ cell around the cell's location $\vJ_k$ with density
proportional to some function $\Blob(\vJ^\star|\vJ_k,\vLambda_k)$,
where $\vLambda_k$ describes the size and shape of the blob.  Then the
DF becomes
\begin{equation}
  f(\vJ)
  =\sum_{k=1}^K\pi_k\Blob(\vJ|\vJ_k,\vLambda_k),
\label{eq:blobbyDF}
\end{equation}
with $\pi_k$ drawn from the Dirichlet
distribution~\eqref{eq:dirichlet} with $\alpha_k=\alpha/K$.  The
parameters $\pi_k$, $\vJ_k$ and $\vLambda_k$ of the blob associated
with each cell are completely independent, save for the fact that
$\sum_k\pi_k=1$.  It is perhaps helpful to think of the
DF~\eqref{eq:blobbyDF} as representing the galaxy by a sum of its
progenitor stellar clusters in phase space, the tidal debris from each
cluster smeared out by two-body encounters and other relaxation
effects, but we emphasise that the blobs are fundamentally purely
formal devices used to introduce neighbouring parts of action space to
one another.

In the absence of any constraints other than the location
parameter~$\vJ_k$ and the scale/shape parameter~$\vLambda_k$, a
natural way of representing each blob would be by using a single
Gaussian.  Recall, however, that at least one of the components of
$\vJ$ is constrained to be non-negative and yet we want $\pi_k$ to be
the total probability mass of the blob.  This means that the function
$\Blob(\vJ|\vJ_k,\vLambda_k)$ must have unit mass when integrated over
the physically allowed region of action space.  To ensure this, we take
\begin{equation}
  \label{eq:blob}
  \Blob(\vJ|\vJ_k,\vLambda_k)=\sum_{m=1}^M{\cal N}(\vJ|R_m\vJ_k,\vLambda_k^{-1}),
\end{equation}
in which ${\cal N}$ is the usual normal distribution,
\begin{equation}
  {\cal N}(\vx|\bar\vx,\vLambda^{-1})
  \equiv\frac{|\vLambda|^{1/2}}{(2\pi)^{d/2}}
  \exp\left[-\frac12(\vx-\bar\vx)^T\vLambda(\vx-\bar\vx)\right],
\label{eq:normal}
\end{equation}
where $|\vLambda|$ is the determinant of the precision (i.e.,
inverse covariance) matrix $\vLambda$,
and $R_1,...,R_M$ are reflection operators that produce mirror images
of the Gaussian at $\vJ=\vJ_k$.  For the case $d=3$ in which
$\vJ=(J_r,J_\theta,J_\phi)$ there are $M=4$ such operators:
\begin{equation}
  \begin{split}
    R_1&=\diag(+1,+1,+1),\\
    R_2&=\diag(-1,+1,+1),\\
    R_3&=\diag(+1,-1,+1),\\
    R_4&=\diag(-1,-1,+1).
  \end{split}
\label{eq:reflection}
\end{equation}
These reflect the original Gaussian centred on $\vJ=\vJ_k$ about the $J_r=0$
and the $J_\theta=0$ axes so that the total mass of the blob in the
physically allowed $J_r\ge0$, $J_\theta\ge0$ subvolume is equal to
one.  Similarly, for the $d=2$ case of $\vJ=(J_r,J_\phi)$ the
necessary reflections are $R_1=\diag(1,1)$, $R_2=\diag(-1,1)$ and for
$d=1$ they are simply $R_1=+1$, $R_2=-1$.
Note that, although the first
argument $\vJ$ of the function $\Blob(\vJ|\vJ_k,\vLambda_k)$ has
restrictions on the signs of some of its components, we do not need to
impose any such restrictions on the second argument, $\vJ_k$.
Therefore we allow the components of $\vJ_k$ to take either
sign, so that, in the three-dimensional case, any of the four choices
$\vJ_k=(\pm J_r,\pm J_\theta,J_\phi)$ refer to the same blob. 

In this scheme each cell has a characteristic width
\begin{equation}
 \Delta J=2J_{\rm box}K^{-1/d} 
\end{equation}
that shrinks as
$K\to\infty$.  A priori, each star has probability
\begin{equation}
  \pr(\vJ_k|\hbox{discrete})=
    1/K,
\end{equation}
of belonging to the blob associated with cell~$k$.  In this point of
view the blobs are treated as being ``pinned'' to the cell locations.
In the continuum limit $K\to\infty$ we can forget about these
underlying cells and use $\pi_k$ and $\vJ_k$ to refer directly to the
probability mass and location of the $k^{\rm th}$ blob, the former
having prior~\eqref{eq:DPdefn} and the latter
\begin{equation}
  \pr(\vJ_k|\hbox{continuous})=
  \begin{cases}
    1/(2J_{\rm box})^d,& \hbox{if all components $|J_{k,i}|<J_{\rm
        box}$},\\
    0,&\hbox{otherwise},
  \end{cases}
\end{equation}
so that $\pr(\vJ_k|\hbox{continuous})(\Delta
J)^d=\pr(\vJ_k|\hbox{discrete})$.  Both ways of thinking about $\vJ_k$
are useful when deriving expressions for the marginal likelihood
(Appendices \ref{sec:exact} and~\ref{sec:VB}).

For the prior $\pr(\vLambda_k)$ on the precision matrix~$\vLambda_k$ we adopt the
uninformative \citep[e.g.,][]{Press12} distribution
\begin{equation}
  \begin{split}
  \pr(\vLambda_k)&=B_0|\vLambda_k|^{-\frac12(d+1)},
  \end{split}
  \label{eq:priorlam1}
\end{equation}
but with a restriction on the range of allowed $\vLambda_k$ to those
that produce blobs that are larger than the cell size $\Delta J$ but
smaller than $J_{\rm box}$.  Appendix~\ref{sec:lambdastuff} gives the
details of how we impose this restriction and calculate the
dimensionless normalisation constant $B_0$.

Given a distribution of stars with actions
$\vJ^\star_1$,...,$\vJ^\star_N$, the likelihood is then
\begin{equation}
  \begin{split}
  \pr(\vJ^\star_1...\vJ^\star_N|\vpi,\{\vJ,\vLambda\})
  &=\prod_{n=1}^N\sum_{k=1}^K\pi_k\Blob(\vJ^\star_n|\vJ_k,\vLambda_k).\\
  \end{split}
  \label{eq:lik1}
\end{equation}
This awkward product of sums can be rewritten as an easier-to-handle
sum of products,
\begin{equation}
  \begin{split}
    \prod_{n=1}^N\sum_{k=1}^K\pi_k\Blob(\vJ^\star_n|\vJ_k,\vLambda_k)
=\sum_{\vZ}  
\prod_{n=1}^N\prod_{k=1}^K\left[\pi_k\Blob(\vJ^\star_n|\vJ_k,\vLambda_k)\right]^{z_{nk}}
  \end{split}
\end{equation}
by introducing a set of binary variables $\vZ=\{z_{nk}\}$ that
indicate which of the $K$ blobs in equation~\eqref{eq:lik1} each of
the $N$ stars comes from: $z_{nk}=1$ if star $n$ comes from the
$k^{\rm th}$ blob and is zero otherwise; for each $n$ there is
precisely one $k$ for which $z_{nk}=1$.  Similarly, as the function
$\Blob(\vJ^\star_n|\vJ_k,\vLambda_k)$ is itself a sum of $M$
Gaussians, we can extend this idea and use $z_{nkm}$ to indicate which
of the $K\times M$ Gaussians each star is drawn from.  Then
$z_{nk}=\sum_m z_{nkm}$.  If we think of $\vZ=\{z_{nkm}\}$ as a
(latent) variable, then the likelihood becomes
\begin{equation}
  \pr(\vJ^\star_1...\vJ^\star_N,\vZ|\vpi,\{\vJ,\vLambda\})
  =  \pr(\vJ^\star_1...\vJ^\star_N|\vZ,\{\vJ,\vLambda\})\pr(\vZ|\vpi),
\label{eq:likZ}
\end{equation}
in which
\begin{equation}
  \begin{split}
  \pr(\vJ^\star_1...\vJ^\star_N|\vZ,\{\vJ,\vLambda\})
  &=\prod_{n=1}^N\prod_{k=1}^K
  \left[\Blob(\vJ^\star_n|\vJ_k,\vLambda_k)\right]^{z_{nk}}\\
  &=\prod_{n=1}^N\prod_{k=1}^K\prod_{m=1}^M
  \left[{\cal N}(\vJ^\star_n|R_m\vJ_k,\vLambda_k^{-1})\right]^{z_{nkm}}
  \label{eq:lik2}
  \end{split}
\end{equation}
and
\begin{equation}
  \label{eq:rprior}
  \pr(\vZ|\vpi) =
  \prod_{n=1}^N\prod_{k=1}^K\pi_k^{z_{nk}}
=\prod_{n=1}^N\prod_{k=1}^K\prod_{m=1}^M\pi_k^{z_{nkm}}.
\end{equation}
The true likelihood~\eqref{eq:lik1} is obtained by
summing~\eqref{eq:likZ} over all possible assignments~$\vZ$ of stars
to blobs (or Gaussians).

Combining this likelihood with the priors on $(\pi,\{\vJ,\vLambda\})$
introduced above yields the ``probability of everything'',
\begin{equation}
  \begin{split}
    &\pr(\vJ^\star_1...\vJ^\star_N,\vZ,\vpi,\{\vJ,\vLambda\})\\
    &\qquad\qquad=\pr(\vJ^\star_1...\vJ^\star_N,\vZ|\vpi,\{\vJ,\vLambda\})
  \pr(\vpi)\pr(\{\vJ,\vLambda\}).
  \end{split}
\label{eq:probofeverything0}
\end{equation}
Since the DF is completely described by the parameters
$(\vpi,\{\vJ,\vLambda\})$, we may obtain the marginal
likelihood~\eqref{eq:sumoverDF} from~\eqref{eq:probofeverything0} by
considering all possible assigments~$\vZ$ of stars to blobs,
marginalising the blob parameters~$(\vpi,\{\vJ,\vLambda\})$ for
each choice of~$\vZ$, and summing the results:
\begin{equation}
  \begin{split}
&\pr(\vJ^\star_1...\vJ^\star_N|\assumptions)\\
&\qquad=\sum_\vZ\int\d\vpi\int\d \vJ_{1...K}\int\d\vLambda_{1...K}\,\pr(\vJ^\star_1...\vJ^\star_N,\vZ,\vpi,\{\vJ,\vLambda\}).
  \end{split}
\label{eq:sumoverblobs}
\end{equation}



\subsection{Summary of probability distributions}
\label{sec:DPMsummary}
The following lists the probability distributions introduced above
and summarises the assumptions~$\assumptions$ made
to calculate the marginal likelihood
$\pr(\vJ^\star_1...\vJ^\star_N|\assumptions)$ of
equation~\eqref{eq:sumoverDF}.  The fundamental assumption is that the
DF is uniform in angle and can be described by a sum of
blobs~\eqref{eq:blobbyDF} in action space.  Then from
equation~\eqref{eq:sumoverblobs} the marginal likelihood
$\pr(\vJ^\star_1...\vJ^\star_N|\assumptions)$ is obtained by
marginalising the ``probability of
everything''
\begin{equation}
  \begin{split}
  \pr(\vJ^\star_1...\vJ^\star_N,\vZ,\vpi,\{\vJ,\vLambda\})
  &=\pr(\vJ^\star_1...\vJ^\star_N|\vZ,\{\vJ,\vLambda\})\pr(\vZ|\vpi)\\
  &\qquad\times\pr(\vpi)\pr(\{\vJ,\vLambda\}),
  \end{split}
\label{eq:probofeverything}
\end{equation}
over the hidden variables $\vZ$ and the model parameters
$(\vpi,(\vJ_1,\vLambda_1),...,(\vJ_K,\vLambda_K))$ in the limit
$K\to\infty$.  The likelihood on the right-hand side
of~\eqref{eq:probofeverything} is has two factors.  One is a product
of Gaussians,
\begin{align}
  \pr(\vJ^\star_1...\vJ^\star_N|\vZ,\{\vJ,\vLambda\})&=\prod_{n=1}^N\prod_{k=1}^K\prod_{m=1}^M
  \left[{\cal N}(\vJ^\star_n|R_m\vJ_{k},\vLambda_k^{-1})\right]^{z_{nkm}},\label{eq:likjstar}
\end{align}
where the reflection operators $R_m$ are given by~\eqref{eq:reflection}.
The other is the multinomial
\begin{equation}
  \pr(\vZ|\vpi)=\prod_{n=1}^N\prod_{k=1}^K\prod_{m=1}^M\pi_k^{z_{nkm}}.
\label{eq:przpi}
\end{equation}
The priors on the model parameters are
\begin{align}
  \pr(\vpi)&=\Dir(\vpi|\valpha_0),\label{eq:priorpi}\\
  \pr(\{\vJ,\vLambda\})&=\prod_{k=1}^K\pr(\vJ_k)\pr(\vLambda_k),\\
  \pr(\vJ_k)&=
  \begin{cases}
    (2J_{\rm box})^{-d}, & \hbox{if all components $|J_{k,i}|<J_{\rm box}$},\\
    0 , & \hbox{otherwise},
  \end{cases}\label{eq:priorj}\\
  \pr(\vLambda_k)&=B_0|\vLambda_k|^{-\frac12(d+1)},\label{eq:priorlam0}
\end{align}
with $\valpha_0=\textstyle\left(\frac \alpha K,...,\frac \alpha K\right)$.
The variables $\alpha$, $J_{\rm box}$ and $B_0$ are (degenerate)
hyperparameters: a brief inspection of equations
\eqref{eq:probofeverything} to \eqref{eq:priorlam0} shows that the
model behaviour is controlled by the single hyperparameter
\begin{equation}
  \alpha'=\frac{\alpha B_0}{(2J_{\rm box})^d},
\label{eq:alphaprime}
\end{equation}
which has dimensions of $(\hbox{action-space volume})^{-1}$.
Following the discussion in Section~\ref{sec:DP}, the larger the value
of $\alpha'$ the larger the prior weight given to clumpy DFs that
are composed of many blobs.

The model defined by equations
\eqref{eq:probofeverything}--\eqref{eq:priorlam0} is a straightforward
variant of the ``infinite mixture of Gaussians'' problem that is well known in
the statistics and machine learning communities \citep[e.g.,][]{Rasmussen00}.
The only differences are that we have introduced the
reflection matrices $R_m$ and, forsaking some computational
convenience, have imposed explicitly noninformative priors on $\vJ_k$
and~$\vLambda_k$.

\subsection{Comparison with the maximum-likelihood
  orbit-superposition (``extended Schwarzschild'') method}

The maximum-likelihood orbit-superposition method
(\S\ref{sec:osmodels}) can be viewed as a special case of the
modelling procedure above in which one chooses a {\it fixed}
set of blob parameters ($\vJ_k,\vLambda_k)$: the locations $\vJ_k$
are set by the choice of orbit library and the precisions are taken to
be $\vLambda_k=\frac1\epsilon I$, where $I$ is the identity matrix and
$\epsilon\to0$, so that each blob contracts to a single orbit.  The
only free parameters are then the ``orbit weights''~$\vpi$, the most
likely set of which is taken to be indicative of all DFs in that
potential.
Section~\ref{sec:osmodels} lists a step-by-step procedure for applying the
maximum-likelihood procedure in practice.  For comparison, here are
the corresponding steps in the Dirichlet process mixture method:
\begin{enumerate}
\item Instead of representing the DF as a weighted sum
  $f(\vJ)=\sum_kw_kf_k(\vJ)$ of fixed basis functions~$f_k(\vJ)$,
  write it as a sum of an arbitrary number of blobs in action
  space~\eqref{eq:blobbyDF}, each blob having some unknown mass $\pi_k$,
  location~$\vJ_k$ and shape~$\vLambda_k$.
\item For each datapoint $(\vx^\star_n,\vv^\star_n)$, calculate the
  contribution to the likelihood,
  \begin{align}
    P_{nk}&=
    \pr((\vx_n^\star,\vv_n^\star)|\vJ_k,\vLambda_k,\Phi)\nonumber\\
    &=\int
  \pr(\vx_n^\star,\vv_n^\star|\vJ_n^\star,\Phi)\Blob(\vJ^\star_n|\vJ_k,\vLambda_k)
  \,\d^d\vJ^\star_n,
  \label{eq:spellitout1}
  \end{align}
  made by an arbitrary blob $(\vJ_k,\vLambda_k)$ in the assumed
  potential~$\Phi$.  For the special case of perfect, unbiased data
  focused on in this paper the
  $\pr(\vx_n^\star,\vv_n^\star|\vJ_n^\star,\Phi)$ factor in the
  integrand is a Dirac delta that picks out the actions $\vJ^\star_n$
  of the orbit that passes through the point
  $(\vx^\star_n,\vv^\star_n)$ and so
  $P_{nk}=\Blob(\vJ_n^\star|\vJ_k,\vLambda_k)$.  Unlike the
  corresponding step in the maximum-likelihood method, at this stage
  we cannot write down a numerical value for $P_{nk}$, because it
  depends on the nuisance parameters $(\vJ_k,\vLambda_k)$ which are
  marginalised in the next step.
\item Perform the marginalisation~\eqref{eq:sumoverblobs}.

\end{enumerate}
Note that the third step above combines the last two steps of
the maximum-likelihood procedure of \S\ref{sec:osmodels} into one.

\section{Tests}
\label{sec:applications}

In this section we present the results of calculating the marginal
likelihood $\pr(D|\Phi)$ given by equations \eqref{eq:PDgivenPhi}
and~\eqref{eq:sumoverblobs} above for some simple test problems,
starting with the one-dimensional simple harmonic oscillator of
Section~\ref{sec:toy}.
We make use of two different schemes to compute 
the $\pr(\vJ^\star_1...\vJ^\star_N|\assumptions)$ given by
equation~\eqref{eq:sumoverblobs}:
\begin{enumerate}
\item an exact calculation obtained by
  reducing~\eqref{eq:sumoverblobs}  to a
  sum over all possible partitions of the set
  of $N$ stars (equ.~\ref{eq:exactmarglike});
\item a variational lower bound (equ.~\ref{eq:simplik}) obtained by
  fitting a simple functional form to the
  integrand~\eqref{eq:probofeverything}.
\end{enumerate}
Although both schemes are simple to implement in practice, their
derivations are quite involved and so we relegate the details to
Appendices \ref{sec:exact} and~\ref{sec:VB} respectively.  Practical
application of the exact calculation~(i) is feasible only for small
numbers of stars, $N<10$.  In contrast, the approximate variational
scheme~(ii) is computationally inexpensive: Appendix~\ref{sec:VBsteps}
provides step-by-step instructions on how to implement it.

\subsection{One-dimensional simple harmonic oscillator}
\label{sec:shodpm}

\begin{figure}
  \centerline{
    \includegraphics[width=0.8\hsize]{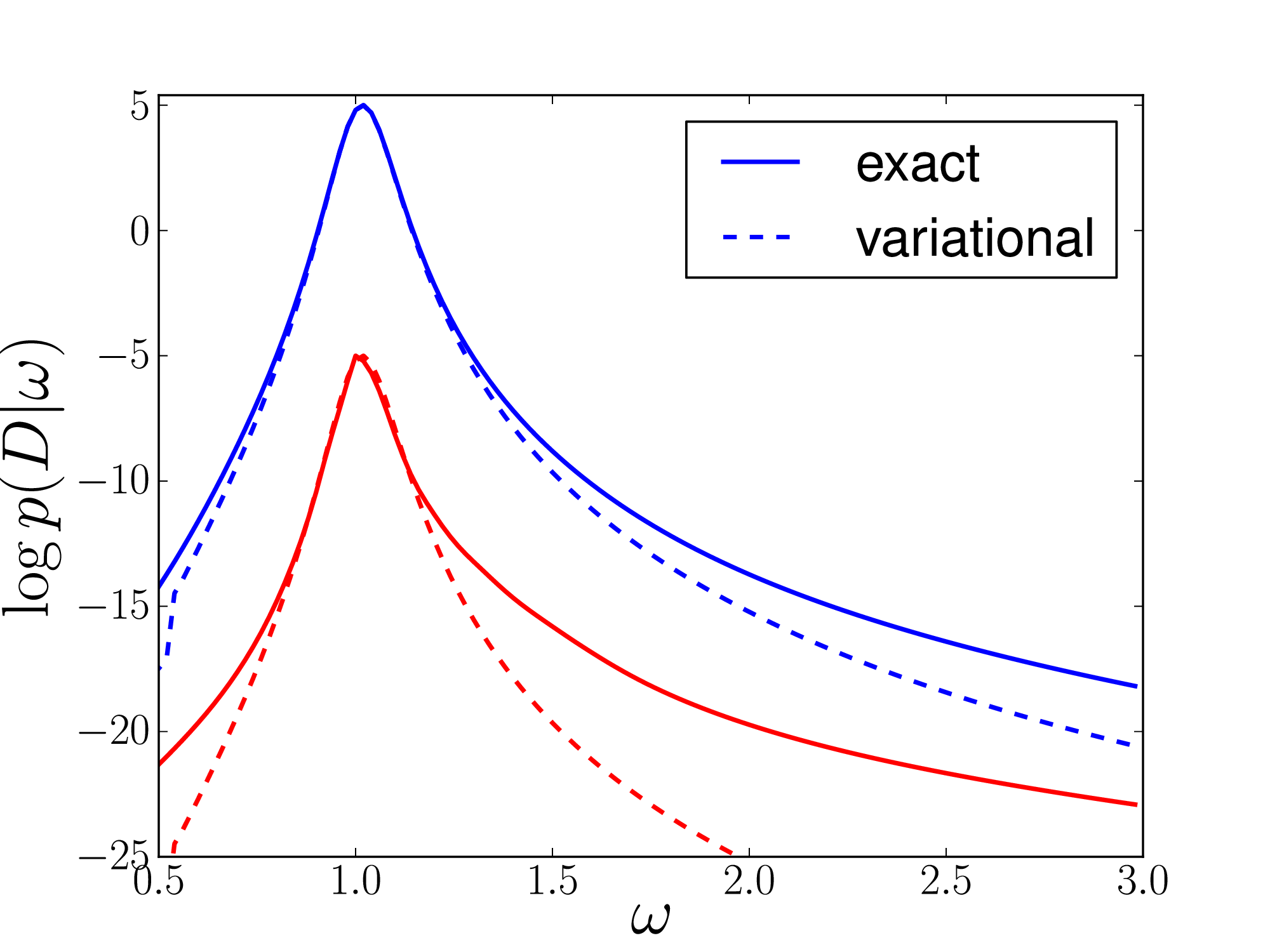}
  }

  \caption{The marginalised likelihood $\pr(D|\omega)$ for the sample
    of 10 stars shown in Figure~\ref{fig:exampleD} calculated using
    the exact method of Appendix~\ref{sec:exact} (solid curves) and
    the approximate variational method of Appendix~\ref{sec:VB}
    (dashed curves).  The upper pair of curves are for concentration
    parameter $\alpha'=10^{-3}$, the lower for $\alpha'=10^{-1}$.  The two pairs of
    curves have been offset vertically from one another for
    clarity. Compare to Figure~\ref{fig:osresults}.}
  \label{fig:shovb}
\end{figure}

In the simple harmonic potential $\Phi(x)=\frac12\omega^2 x^2$ the
action associated with a star that passes through the point
$(x,v)$ is 
\begin{equation}
  J(x,v|\omega)=\frac1{2\pi}\oint \dot x\,\d x=
\frac{\omega}{2\pi}\left(x^2+\frac{v^2}{\omega^2}\right).
\end{equation}
For each of the sample~$D$ of $N=10$ stars shown in
Figure~\ref{fig:exampleD} changing the assumed $\omega$ changes the
action $J_n^\star=J(x^\star_n,v^\star_n|\omega)$.
Figure~\ref{fig:shovb} plots the marginal likelihood
$\pr(D|\omega)=\pr(J_1^\star,...,J_N^{\star})$ for a range of assumed
values of $\omega$ using both the exact calculation of
Appendix~\ref{sec:exact} and the variational estimate of
Appendix~\ref{sec:VB}.

This plot demonstrates two important points.
The first, more practical, point is that the variational estimate
(Appendix~\ref{sec:VB}) of the marginal likelihood agrees well with
the exact calculation (Appendix~\ref{sec:exact}), especially for small
values of the concentration parameter~$\alpha'$
(equ.~\ref{eq:alphaprime}).  This is not surprising, as the
variational estimate essentially approximates the
integrand~\eqref{eq:probofeverything} by the contribution made by one
or more well-chosen blobs, which will tend to be a good approximation
for small~$\alpha$.  This is important, because even for only $N=10$
stars the exact calculation of $\pr(D|\omega)$ takes a few seconds per
$\omega$ on a standard PC, whereas the variational approximation is
instantaneous.

\begin{figure*}
  \centerline{
    \includegraphics[width=0.4\hsize]{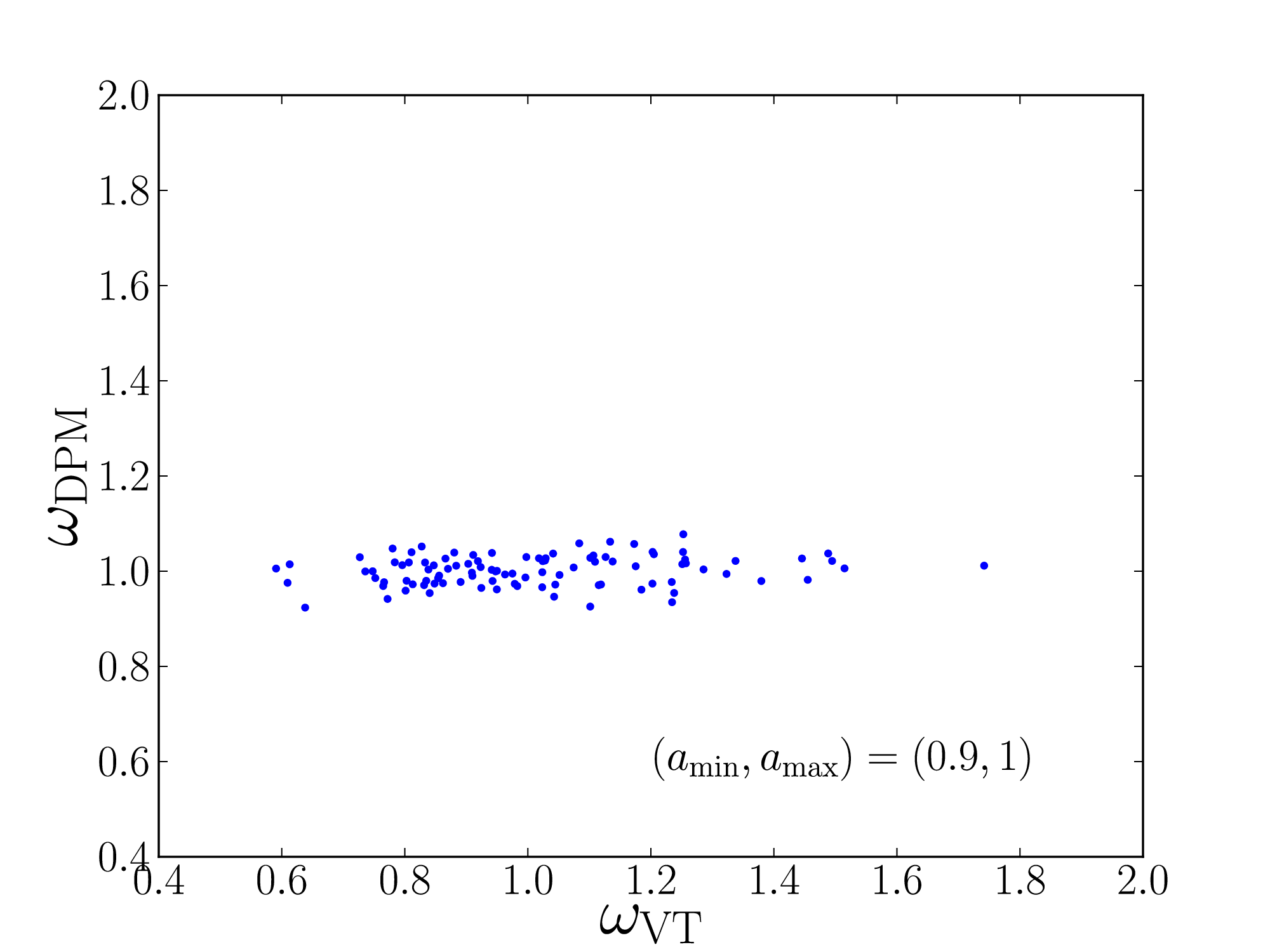}
    \includegraphics[width=0.4\hsize]{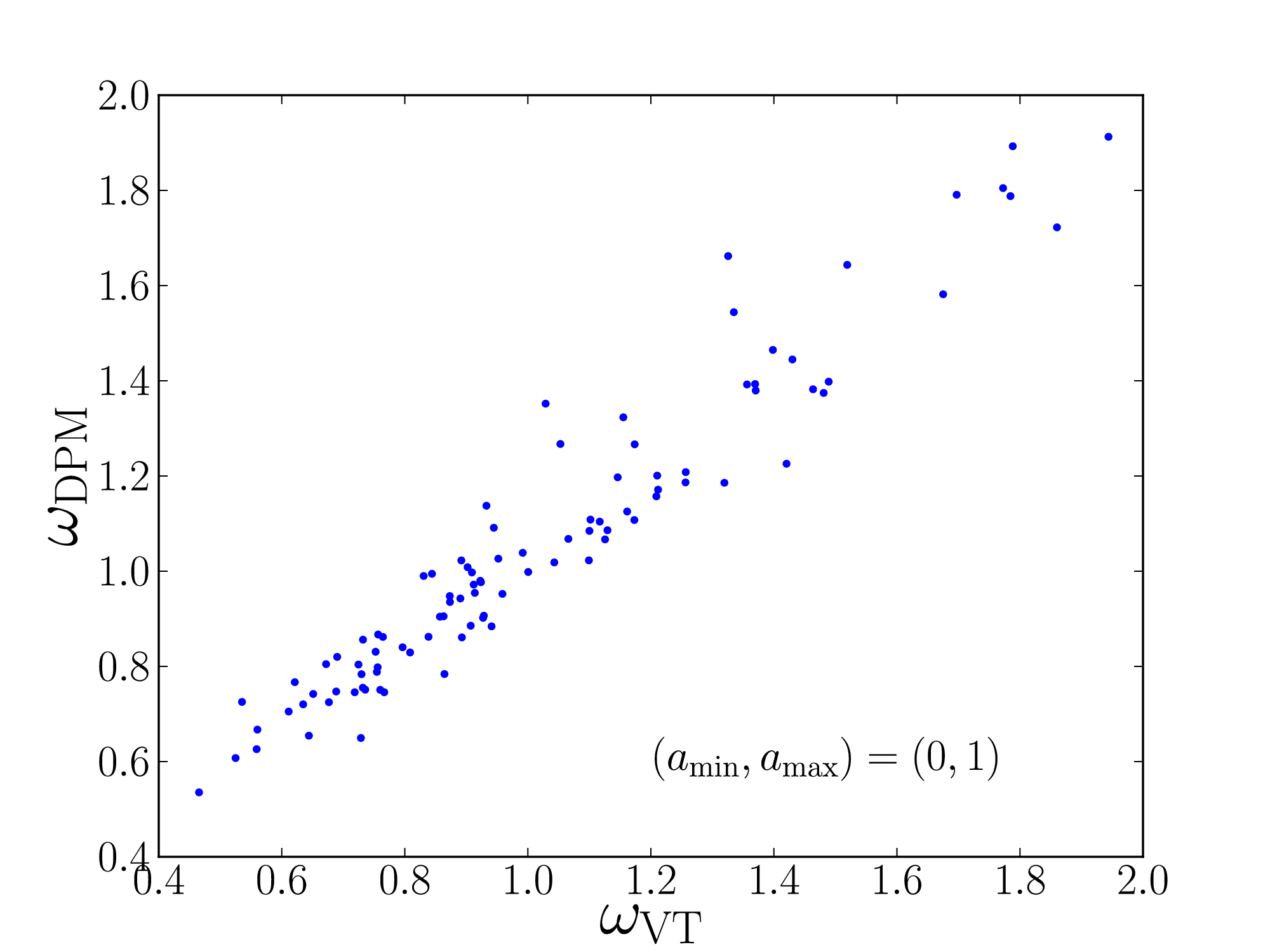}
  }

  \caption{Comparison of the virial theorem estimate $\omega_{\rm VT}$
    (equ.~\ref{eq:VT2}) to the marginal likelihood-based estimate $\omega_{\rm DPM}$
    (equ.~\ref{eq:omegadpm}) 
    for many realisations of a simple harmonic oscillator model
    having $N=10$ stars distributed uniformly in amplitude~$a$ between
    $a_{\rm min}$ and $a_{\rm max}$.  The panel on the left plots the
    comparison for $(a_{\rm min},a_{\rm max})=(0.9,1)$.  The one on
    the right is for $(0,1)$.}
  \label{fig:shorealisations}
\end{figure*}

The second point to note from Figure~\ref{fig:shovb} is that, despite
the relatively small number of stars, the marginal likelihood is
sharply peaked about the correct value of $\omega=1$.  The reason for
this is that the Dirichlet process mixture used to model the DF
favours DFs that are strongly peaked in action space.  A quantitative
explanation of this is given at the end of Appendix~\ref{sec:exact},
but to illustrate it we have constructed realisations of two different
simple-harmonic oscillator models.  The models have $\omega=1$ and
$N=10$ stars drawn from a uniform distribution of stellar amplitudes
$a$ between some $a_{\rm min}$ and $a_{\rm max}$.  For one model we
take the same narrow distribution of amplitudes, $(a_{\rm min},a_{\rm
  max})=(0.9,1.0)$, as used in Figures~\ref{fig:exampleD}
to~\ref{fig:shovb}.  For the other we use the broader $(a_{\rm
  min},a_{\rm max})=(0,1)$.  For each realisation of each model
Figure~\ref{fig:shorealisations} compares the virial theorem
estimate~\eqref{eq:VT2} of $\omega$ to the expectation value,
\begin{equation}
  \omega_{\rm DPM}=\int\pr(D|\omega)\pr(\omega)\d\omega,
\label{eq:omegadpm}
\end{equation}
obtained from the marginal likelihood~$\pr(D|\omega)$
(equ.~\ref{eq:PDgivenPhi}) with an uninformative prior
$\pr(\omega)\propto1/\omega$.  It is evident from
Figure~\ref{fig:shorealisations} that the estimate~\eqref{eq:omegadpm}
is much better than $\omega_{\rm VT}$ in the case of the model with
the narrow distribution of amplitudes: $\omega_{\rm DPM}$ is always
much closer the correct $\omega=1$ than $\omega_{\rm VT}$.    The
two estimates are comparable when the amplitude distribution is broad, however.


\begin{figure*}
  \centerline{
    \includegraphics[width=0.3\hsize]{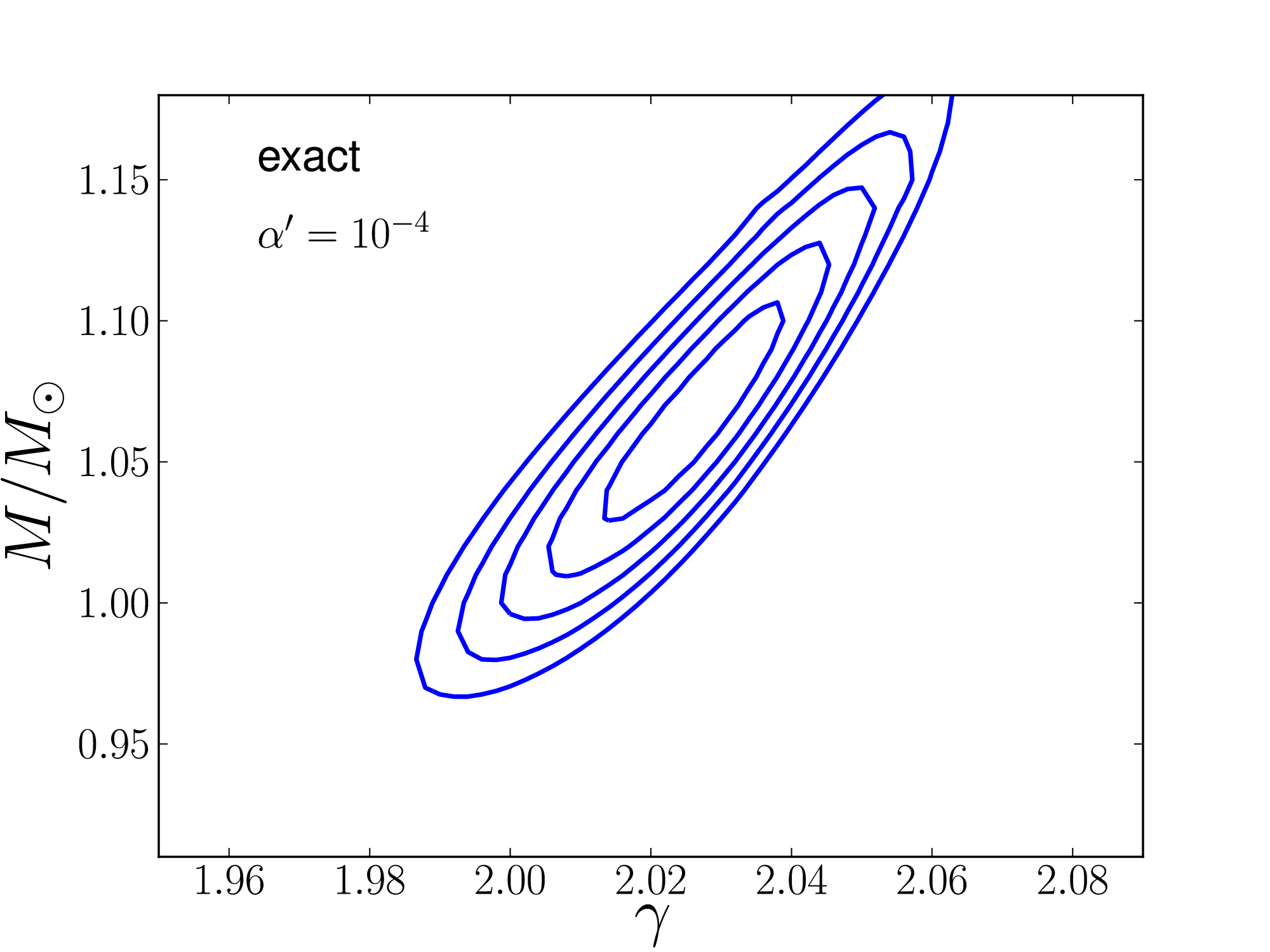}
    \includegraphics[width=0.3\hsize]{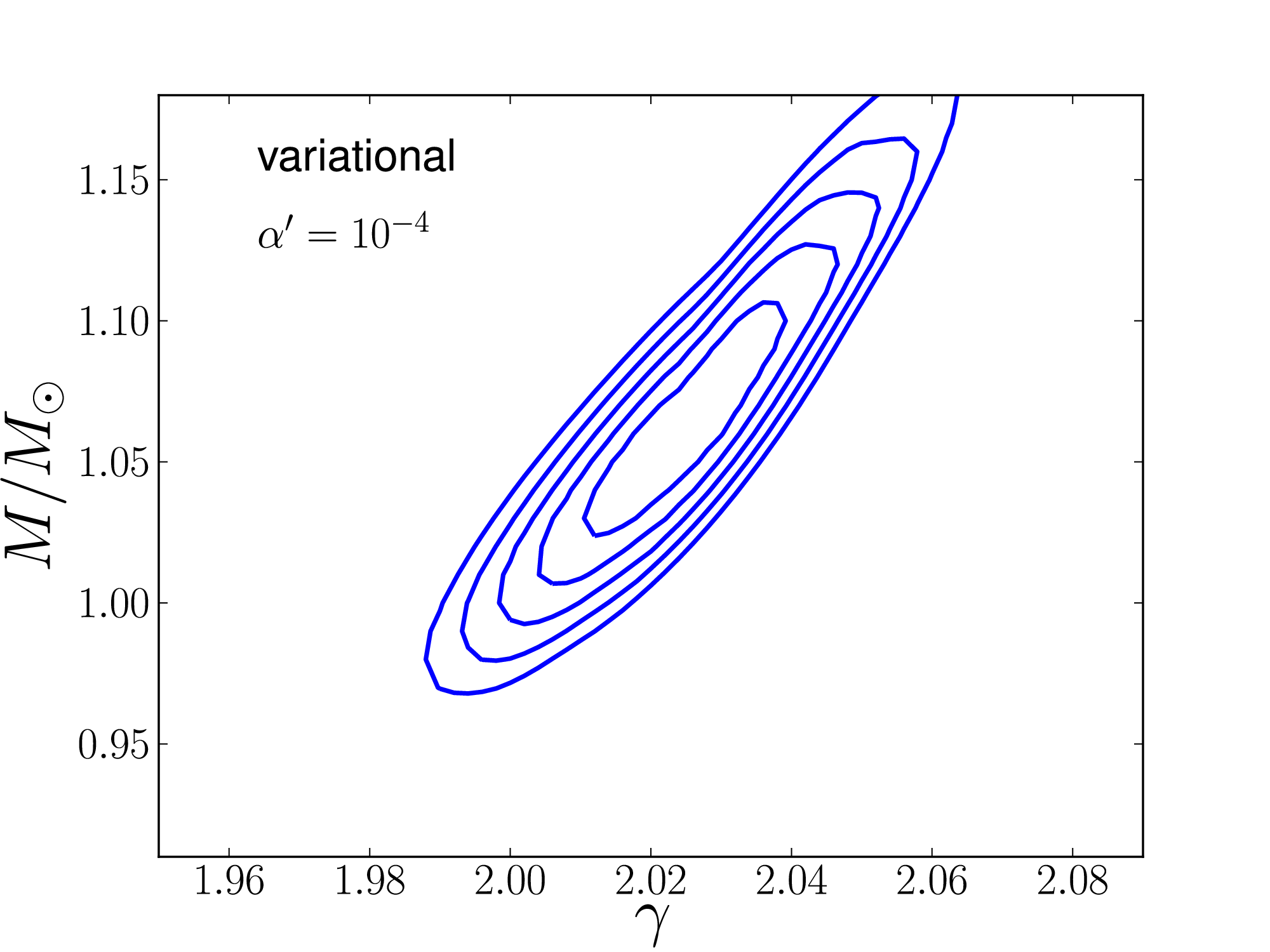}
    \includegraphics[width=0.3\hsize]{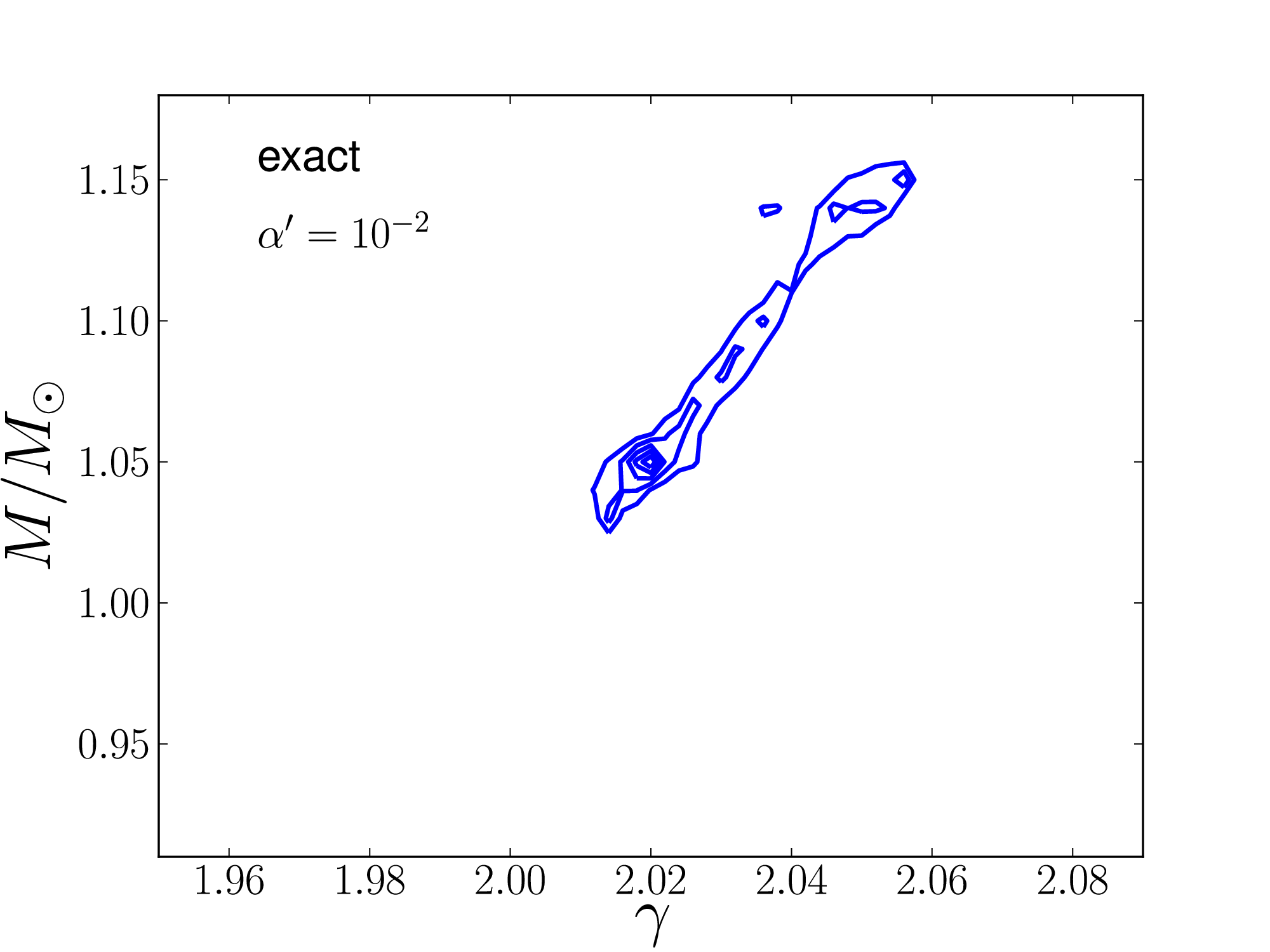}
}
  \caption{Plots of $\log\pr(D|\gamma,M)$ for the solar system problem
    of Section~\ref{sec:ss}.  The panel on the left shows
    the results of the exact calculation with 
    (Appendix~\ref{sec:exact}) for concentration parameter
    $\alpha'=10^{-4}\,\hbox{yr (AU)}^{-2}$.
    The middle panel shows the corresponding variational
    estimate (Appendix~\ref{sec:VB}).
    The panel on the right plots the exact result for 
    $\alpha'=10^{-2}\,\hbox{yr (AU)}^{-2}$.
  }
  \label{fig:ss}
\end{figure*}
\subsection{Solar system}
\label{sec:ss}
\cite{Bovy+10} have recently applied a broadly similar
technique to constrain the force law in the solar system using only a
snapshot of the positions and velocities of the eight major planets at
a specific instance in time, albeit by assuming a parametric form for
the DF from which the planets are drawn.  
We follow them by assuming
that the potential in the solar system is of the form
\begin{equation}
  \Phi(R) = -\frac{GM}{(\gamma+1)R_0}\left(\frac{R_0}R\right)^{\gamma+1},
\end{equation}
so that a planet at radius~$R$ feels a radial acceleration of magnitude
\begin{equation}
  |\ddot{\vx}|=-\frac{GM}{R_0^2}\left(\frac {R_0}R\right)^\gamma.
\end{equation}
We set the reference radius $R_0=1\,\hbox{AU}$, leaving $\gamma$
and $M$ as free parameters.  To illustrate the application of the
method of Section~\ref{sec:solution} to a system with $d=2$ dimensions
we ignore the motion out of the ecliptic plane.  Then for an assumed
$(\gamma,M)$, the two actions associated with an orbit that passes through
the point $(x,y)$ in the ecliptic plane with velocity $(v_x,v_y)$ are
\begin{equation}
  \begin{split}
  J_R(\vx,\vv|\Phi)&=\frac1{2\pi}\oint v_R\,\d R
  =\frac1\pi\int_{R_-}^{R_+}\left[2(E-\Phi)-\frac{L^2}{R^2}\right]^{1/2}
  \,\d R,\\
  J_\phi(\vx,\vv|\Phi) & =\frac1{2\pi} \oint p_\phi\,\d\phi = L,
  \end{split}
\end{equation}
where $L=xv_y-yv_x$ and $E=\frac12\vv^2+\Phi$ are the angular
momentum and energy per unit mass respectively, and $R_\pm(E,L)$ are
the corresponding apo- and peri-helion radii.

For each assumed $(\gamma,M)$ we calculate the actions
$(J^\star_{R,n},J^\star_{\phi,n})$ associated with the positions
$(x^\star_n,y^\star_n)$ and velocities $(v^\star_{x,n},v^\star_{y,n})$
of the eight major planets from the 1 April 2009 ephemeris in Table
1 of \cite{Bovy+10}.  Figure~\ref{fig:ss} shows the marginalised
likelihood
$\pr(D|\gamma,M)=\pr(\{(J^\star_{R,n},J^\star_{\phi,n})\}|\assumptions)$
calculated using both the exact method of Appendix~\ref{sec:exact} and
the variational estimate of Appendix~\ref{sec:VB}.  As in the case of
the simple harmonic oscillator, the two methods agree well when the
concentration parameter $\alpha'$ is small.  For larger~$\alpha'$ the
exact $\pr(D|\Phi)$ becomes implausibly clumpy, which cannot be
reproduced by the variational estimate.

The resulting $\pr(D|\gamma,M)$ is broadly similar to that obtained by
\citet[][their Figure 6]{Bovy+10}, who assumed various parametrised
forms for the DF and used a Markov-Chain Monte Carlo method to explore
the posterior distribution of their DF and potential parameters
simultaneously.  There are some differences: we see no evidence of the
multimodal structure they found, and our posterior probability
distribution is slightly tighter, with a stronger covariance between
$\gamma$ and~$M$.  In common with them, we find that the model is only
marginally consistent with the correct result of
$(\gamma,M)=(2,M_\odot)$.  The marginal likelihood peaks at
$(\gamma,M)\simeq(2.02,1.07\,M_\odot)$: the model slightly
overestimates the inward acceleration felt by the inner planets
(including the earth), and slightly underestimates the accelerations
further out.  On the other hand, if we believe from Poisson's equation
that $\nabla^2\Phi\ge0$ everywhere, then we must impose the prior
constraint that $\gamma\le2$ and the resulting $\pr(D|\gamma,M)$ is
strongly peaked very close to the correct $M=M_\odot$ value.

\subsection{A simple galaxy model}
\label{sec:galaxy}
A more realistic example is provided by a catalogue of stars taken
from a snapshot of an equilibrium galaxy model.  The toy galaxy has a
black hole of mass $M_\bullet$ embedded in a 
uniform, spherical distribution of dark matter with mass $M_0$ within
a reference radius $r_0$.  The potential is then
\begin{equation}
  \label{eq:galpot}
  \Phi(r)=-\frac{GM_\bullet}r+\frac{GM_0}{2r_0^3}r^2.
\end{equation}
Throughout the following we set the reference radius $r_0=1$.  The
stars in the galaxy are luminous test particles with a \cite{Hernquist90}
number-density profile,
\begin{equation}
  \label{eq:hernq}
  \rho(r)\propto\frac1{r(r_{\rm H}+r)^3}
\end{equation}
and an isotropic internal velocity distribution.   To create the
simulated catalogue we use Eddington's
formula to find the distribution function $f(E)$ that produces the
number-density profile~\eqref{eq:hernq} in the potential~\eqref{eq:galpot}
and then generate mock observations by drawing
$(\vx^\star_n,\vv^\star_n)$ for $n=1,...,N=10^4$ stars from this DF.
The model galaxy has $M_\bullet=M_0=1$, so that $r_0$ is approximately
equal to the radius of the sphere of influence of the black hole.  We
choose $r_{\rm H}=r_0/(1+\sqrt2)$, which places half of the stars
inside~$r_0$.

\begin{figure}
  \centerline{\includegraphics[width=0.8\hsize]{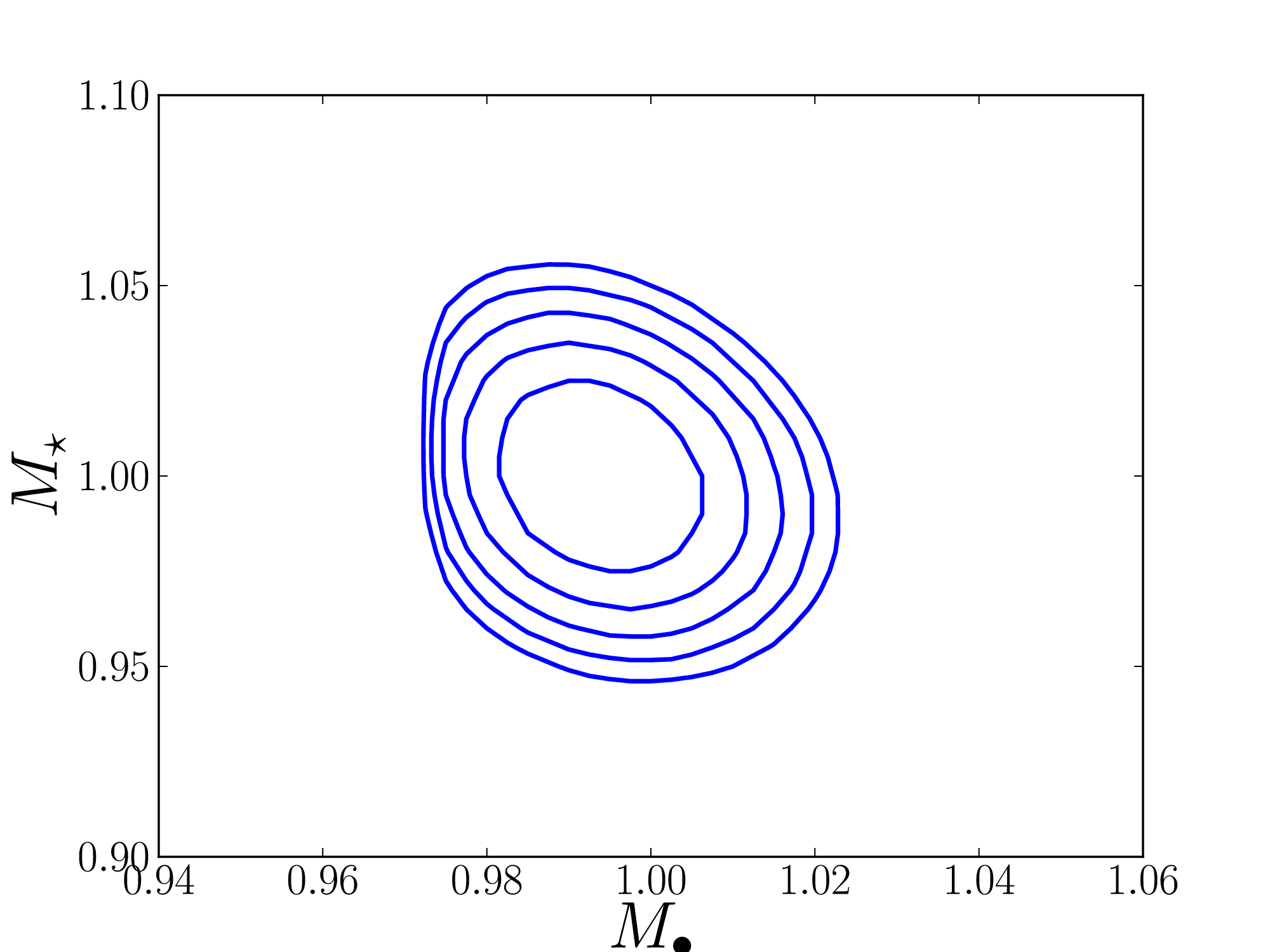}}

  \caption{Plot of $\log\pr(D|M_\bullet,M_\star)$ for the toy galaxy
    model of Section~\ref{sec:galaxy}.
    Successive contour levels are spaced at
    $\Delta\log\pr(D|M_\bullet,M_\star)=1$.
  }
  \label{fig:gal}
\end{figure}

Having this catalogue of $10^4$ stars we use the variational method of
Appendix~\ref{sec:VB} to approximate the marginal likelihood
$\pr(D|\Phi)$ for potentials $\Phi(r)$ of the form~\eqref{eq:galpot}
with different assumed $(M_\bullet,M_\star)$.  For each assumed
$(M_\bullet,M_\star)$, the actions associated with an orbit that passes
through the phase-space point $(\vx,\vv)$ are
\begin{equation}
  \begin{split}
  J_r(\vx,\vv|\Phi)&=\frac1{2\pi}\oint v_r\,\d r
  =\frac1\pi\int_{r_-}^{r_+}\left[2(E-\Phi)-\frac{L^2}{r^2}\right]^{1/2}
  \,\d r,\\
  J_\phi(\vx,\vv|\Phi) & =\frac1{2\pi} \oint p_\phi\,\d\phi = L_\phi,\\
  J_\theta(\vx,\vv|\Phi) &= L-|L_\phi|,
  \end{split}
\end{equation}
where $L=|\vx\times\vv|$, $L_\phi=xv_y-yv_x$ and $E=\frac12\vv^2+\Phi$
are the total angular momentum, its projection onto the $z$ axis and
the energy per unit mass respectively, and $r_\pm(E,L)$ are the apo-
and peri-centre radii.\footnote{This $f(J_r,J_\theta,J_\phi)$ form
  assumed for the DF means that all axisymmetric-in-$z$ DFs are
  included in the marginalisation.  To restrict to spherically
  symmetric distributions we should take $f=f(J_r,J)$, where
  $J=J_\theta+J_\phi$.  The most general DF in spherically symmetric
  potential is a function of {\it four} isolating integrals of motion
  (e.g., $J_r,J_\theta,J_\phi$ and $\Omega,$ the longitude of the ascending
  node); only two of the angle variables need be uniformly distributed.
  There is nothing fundamentally different about applying the method
  presented here to such DFs.
}
  On an unremarkable standard PC the time taken
to do the conversion from $\{(\vx^\star_n,\vv^\star_n)\}$ to
$\{\vJ_n^\star\}$ for the full sample of $N=10^4$ stars and construct
the variational estimate was about 1 second for each
$(M_\bullet,M_\star)$.  The resulting marginalised likelihood
(Figure~\ref{fig:gal}) peaks very close to the correct
$(M_\bullet,M_\star)=(1,1)$ parameters from which the stars were
drawn.

\section{Discussion}
\label{sec:discuss}

\subsection{The choice of prior}

The modelling framework proposed in the present paper is a development
of the ideas set out in \cite[][hereafter M06]{magog06}.  As in M06,
the likelihood $\pr(D|\Phi)$ is obtained by marginalising the DF $f$
from the joint likelihood $\pr(D|\Phi,f)$ with a suitably chosen
prior.  One way of setting this prior would be by imposing a
functional form for the DF that is described by a handful of
parameters.  Then $\pr(D|\Phi)$ is obtained by marginalising these
parameters \citep[e.g.,][]{Ting+12,McMillanBinney13}.  Ideally,
however, one would like to make as few asumptions as possible about
the form of the DF, which raises the question of how to define a
sensible prior on the set of all possible DFs.

\subsubsection{The infinite-divisibility condition}

Both M06 and the present paper use a flexible ``non-parametric'' model
for the DF, which is described by an (infinite) list of prior weights
$\pi_k$.  For simiplicity, these are assumed to be independent of one
another.  For consistency, they are required to satisfy the
agglomerative condition~\eqref{eq:agglom}.  Taken together, these two
assumptions meant that the prior is infinitely divisible.

The most significant difference between the present paper and M06 is
the introduction of blobs.  In M06 the $\pi_k$ gave the DF directly:
each $\pi_k$ was the probability of finding a star within the (tiny)
phase-space volume occupied by the $k^{\rm th}$ cell.  In the present
paper, blobs are introduced to smear out the probability mass $\pi_k$
over neighbouring regions of action space: the DF at any point is the
superposition of overlapping contributions from many smeared-out
cells.
As commented on in both M06 and in Section~\ref{sec:DP} of the present
paper, imposing the infinite-divisibility condition on the DF itself
results in spiky DFs, leading to a flat marginalised likelihood
$\pr(D|\Phi)$ when the data are too good.  The blobs are therefore
essential.  That they have the Gaussian form assumed in this paper is
not.  Nevertheless, the Gaussian assumption is both convenient and
plausible.

M06 was able to evade the blobs by considering only the
problem of calculating $\pr(D|\Phi)$ when the data $D$ were
realistically noisy, integrated line-of-sight velocity distributions.
This leads to a complicated joint likelihood function $\pr(D|\Phi,f)$
that introduces strong correlations among different subvolumes of
phase space.  In contrast, the present paper tackles the problem of
estimating the potential from a discrete, unbiased, error-free sample
of the DF, the joint likelihood $\pr(D|\Phi,f)$ of which introduces no
coupling whatsoever between different regions of phase space, apart
from those required by the strong Jeans theorem.

The other difference between M06 and the present paper is the specific
choice of prior.
Let $V_l$ be a volume of action space and $F(V_l)\equiv\sum_{V_k\in
  V_l}\pi_k$ be the enclosed probability mass before convolution with
the blobs.  The
infinite-divisibility criterion (equ~\ref{eq:agglom} together with the
assumption that the $F_i$ are independent) means that the Laplace
transform of the prior on~$F$,
\begin{equation}
  \bar\pr(s|\alpha)\equiv\int_0^\infty \d F\,\e^{-sF}\pr(F|\alpha),
\end{equation}
must be of the form \citep{Feller66}
\begin{equation}
  \bar\pr(s|\alpha)=\exp\left[-\int_0^\infty\frac{1-\e^{-sF}}{F}
{\cal M}(\alpha,\d F)\right],
\end{equation}
which is completely controlled by the choice of $\alpha$ and the
L\'evy measure ${\cal M}(\alpha,\d F)$.  In M06 we used the galaxy's
luminosity profile as additional prior information on the DF and took
${\cal M}(\alpha,\d F)=\alpha F\e^{-F}\d F$, which can be regarded as
the least informative choice given such extra information on the
expectation values of $F$ \citep{Skilling98}.  In the present paper we
avoid using any such additional information and adopt the
uninformative ${\cal M}(\alpha,\d F)=\alpha\e^{-F}\,\d F$, which
results in the Dirichlet prior~\eqref{eq:dirichlet}.

\subsubsection{Alternative choices for the prior}

The discussion above makes it clear that requiring some form of
correlation among the cells representing the DF is essential if we are
to have any hope of distinguishing one potential from another.  An
alternative model for the prior on the DF would be to drop the
requirement that the $\pi_k$ are independent and satisfy the
agglomerative condition~\eqref{eq:agglom}, imposing instead the weaker
condition that they are drawn from a random process.  For example, one
could sample DFs from a logistic normal process
\cite[e.g.,][]{Lenk88}.  The assumed mean and covariance functions,
$\mu(\vJ)$ and $\sigma^2(\vJ_1,\vJ_2)$ in this process take over the
role of the hyperparameter $\alpha$ and the blobs.  \cite{Bovy+10}
describe some experiments in this area in their Section~6.2 (see also
their Figure~9).

\subsection{Actions versus angles}
In direct contrast to models constructed using the maximum penalised
likelihood method \citep[e.g.,][]{Merritt93}, the models presented
here prefer potentials in which the DF develops sharp features in
action space: broadly speaking, the sharper the DF, the larger the
value of the marginalised likelihood $\pr(D|\Phi)$ (see Section~4 and
Appendix~B).
This is qualitatively similar to \citeapos{PenarrubiaKoposovWalker12}
scheme for constraining the Galactic potential from tidal
streams by looking for potentials that {\it minimise} some estimate of
the entropy of the stars in the stream.  It is instructive to consider
why such a ``minimum-entropy'' method might work.  Their scheme estimated
the entropy based only on the energy of the stars' orbits, but in the
following we generalise it to use actions.  The true entropy
\begin{equation}
  \label{eq:entropy}
  \begin{split}
  S[f]&=-\int f(\vx,\vv)\log f(\vx,\vv)\,\d^d\vx\d^d\vv\\
  &= -\int f(\vJ,\vtheta)\log f(\vJ,\vtheta)\,\d^d\vJ\d^d\vtheta,
  \end{split}
\end{equation}
is actually independent of the potential: it would be futile to use this $S$
to constrain~$\Phi$.
Instead, Pe\~narrubia et al's scheme is equivalent to taking an
orbit-averaged DF,
\begin{equation}
 \bar f(\vJ)=(2\pi)^{-d}\int
f(\vJ,\vtheta)\d^d\vtheta,
\end{equation}
which depends on the assumed~$\Phi$, and looking at how the
orbit-averaged $S[\bar f]$ varies as~$\Phi$ changes.  When the correct
$\Phi$ is used, we have that $\bar f=f$ and so $S[\bar f]=S[f]$.  As
the assumed $\Phi$ moves further from the correct one, the
distribution of angles becomes less uniform, but the true entropy
$S[f]$ remains unchanged.  Therefore $S[\bar f]$ must increase
\footnote{An alternative way of showing this is by expressing $f$ in
  equation~\eqref{eq:entropy} as the Fourier series
  $f(\vJ,\vtheta)=\sum_{\vn}f_\vn(\vJ)\e^{{\rm i}\vn\cdot\vtheta}$ in
  which $f_{-\vn}(\vJ)=f_{\vn}^\star(\vJ)$ and then Taylor expanding
  the logarithm in the integrand about $f_0(\vJ)=\bar f(\vJ)$.  } and
potentials that minimise this $S[\bar f]$ are most consistent with
having a flat distribution in angle.

An alternative way of constraining $\Phi$ would be to try to measure
directly how much the distribution of angles~$\vtheta$ deviates from a
uniform distribution: this is the basis of the ``orbital roulette''
idea proposed by \cite{BeloborodovLevin04}.  It is unclear how to
construct a suitable direct test of non-uniformity though.
There are, however, some very special cases for which one can use an
alternative method to obtain $\pr(D|\Phi)$ directly from the angle
distribution.  For example, in the one-dimensional simple harmonic
oscillator problem\footnote{I thank Scott Tremaine (private
  communication) for pointing this out.}, the quantity $\nu\equiv
v/x$ depends only the angle coordinate~$\theta$.  It is easy to show
that a star observed with $\nu^\star=v^\star/x^\star$ contributes a
factor
\begin{equation}
  \begin{split}
    \pr(\nu^\star|\omega)&=\pr(\theta^\star)\frac{\d\theta^\star}{\d\nu^\star}
    =\frac1{2\pi}\frac{\omega}{\omega^2+{\nu^\star}^2}\\
    &=\frac1{2\pi}\frac{\omega {x^\star}^2}{\omega^2{x^{\star}}^2+{v^\star}^2}
  \end{split}
\end{equation}
to the likelihood of the parameter~$\omega$ that appears in the
potential, independent of any assumptions about $f(J)$.

\begin{figure}
  \centering
  \includegraphics[width=\hsize]{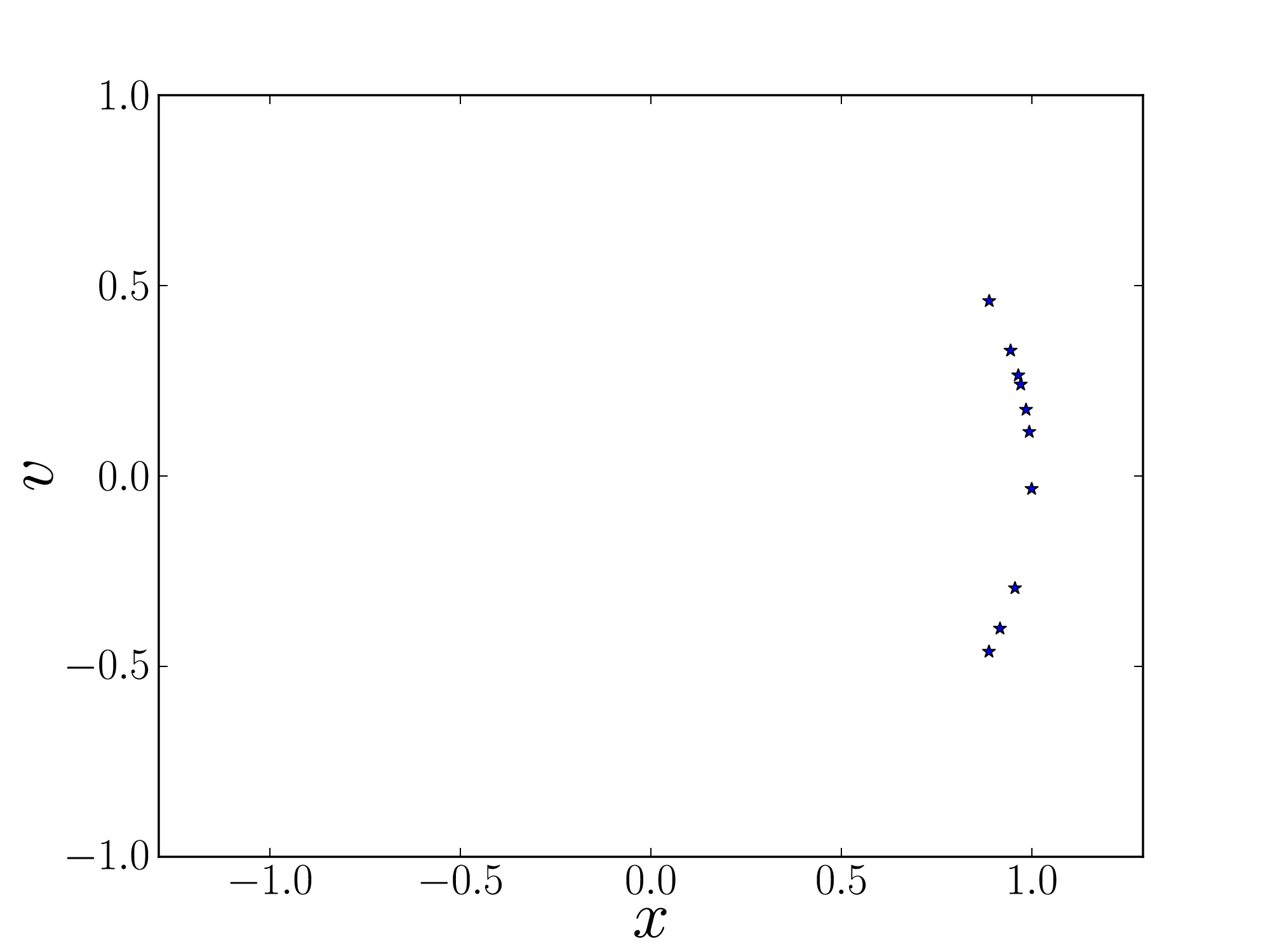}

  \caption{A contrived sample of stars drawn from a one-dimensional
    galaxy model with potential $\Phi(x)=\frac12\omega^2x^2$ with
    $\omega=1$.  The $\pr(D|\omega)$ calculated using the Dirichlet
    process mixture scheme presented in this paper peaks very strongly
    at $\omega=1$, for which the action-space distribution becomes
    sharpest.  On the other hand, the distribution of the stars in
    {\it angle} is flattest when $\omega$ is far from~1. }

  \label{fig:semicirc}
\end{figure}

In common with the minimum-entropy idea above, the Dirichlet process
mixture scheme presented in this paper does not examine the angle
distribution directly.  Instead, it assumes from the outset that the
angle distribution is uniform, so that the DF $f=f(\vJ)$.  This
$f(\vJ)$ becomes sharper as neighbouring tori become more densely
populated when the assumed $\Phi$ tends to the correct one.  The two
approaches -- examining the $\vJ$ distribution versus examining the
$\vtheta$ one -- are not equivalent, as can be seen by considering the
case of a distribution of stars that is incompletely phase mixed
(Figure~\ref{fig:semicirc}).

\section{Generalisation}
\label{sec:gener}

\subsection{Observational errors and selection effects}

\def\obs{^{\rm o}} \def\obs{} Real catalogues suffer from complicated
selection biases and large, correlated errors on the measurements of
individual stars.  Neither of these are difficult to model, at least
in principle.  Suppose that the data $D$ are a list of the stars' {\it
  observed} positions and velocities
$(\vx\obs_1,\vv\obs_1),\ldots,(\vx\obs_N,\vv\obs_N)$, some of which
(e.g., the position of each star along the line of sight or the
projection of its velocity onto the plane of the sky) might have huge
uncertainties.  As before, let us use $(\vx^\star_n,\vv^\star_n)$ to
mean the $n^{\rm th}$ star's {\it true} position and velocity, and
$(\vJ^\star_n,\vtheta^\star_n)$ the corresponding action-angle
coordinates for an assumed trial potential~$\Phi$.  The marginal
likelihood~\eqref{eq:PDgivenPhi} becomes
\begin{equation}
\label{eq:PDgivenPhi2}
\begin{split}
\pr(D|\Phi,\assumptions)
&=\int\d^d\vJ^\star_1\d^d\vtheta^\star_1
   \pr(\vx\obs_1,\vv\obs_1|\vJ_1^\star,\vtheta^\star_1,\Phi)
\cdots\\
&\int\d^d\vJ^\star_N\d^d\vtheta^\star_N
  \pr(\vx\obs_N,\vv\obs_N|\vJ_N^\star,\vtheta^\star_N,\Phi)
\,\cdot\,
\pr(\vJ_1^\star\cdots\vJ_N^\star|\assumptions),
\end{split}
\end{equation}
in which
\begin{equation}
  \begin{split}
    &\pr(\vx\obs,\vv\obs|\vJ^\star,\vtheta^\star,\Phi)\\
    &\quad=\int\d^d\vx^\star\d^d\vv^\star
  \pr(\vx\obs,\vv\obs|\vx^\star,\vv^\star)
  \pr(\vx^\star,\vv^\star|\vJ^\star,\vtheta^\star,\Phi),
  \end{split}
\end{equation}
with the observational uncertainties entering through the factors
$\pr(\vx\obs,\vv\obs|\vx^\star,\vv^\star)$ that relate the observed
position and velocity of each star $(\vx\obs,\vv\obs)$ to their true
values $(\vx^\star,\vv^\star)$.
With this change, the probability of
everything~\eqref{eq:probofeverything} can be written as
\begin{equation}
\begin{split}
&\pr(D,Z,\vpi,\{\vJ,\vLambda\}|\Phi)\\
&\quad=\left[\prod_{n=1}^N\pr(\vx\obs_n,\vv\obs_n,\vZ_n|\vpi,\{\vJ,\vLambda\},\Phi)\right]
\pr(\vpi)\pr(\{\vJ,\vLambda\}),
\end{split}
\end{equation}
where
\begin{equation}
  \begin{split}
&\pr(\vx\obs_n,\vv\obs_n,\vZ_n|\vpi,\{\vJ,\vLambda\},\Phi)\\
&\quad=  \int\d^d\vJ^\star_n\d^d\vtheta^\star_n\,
  \pr(\vx\obs_n,\vv\obs_n|\vJ^\star_n,\vtheta^\star_n,\Phi)
  \pr(\vJ^\star_n,\vZ_n|\vpi,\{\vJ,\vLambda\})
  \end{split}
\label{eq:starbystarlik}
\end{equation}
is the $n^{\rm th}$ star's contribution to the likelihood
$\pr(D,\vZ|\vpi,\{\vJ,\vLambda\})$.

Until now we have assumed that the catalogue~$D$ is an unbiased sample
of the galaxy's underlying DF~$f$.  In reality, catalogues cannot be
unbiased; there are usually constraints on, e.g., the apparent
magnitudes of the stars that are included, or their positions on the
sky.  Any modelling scheme must take such selection effects into
account.  As a simple example of how to model selection effects in the
present scheme, suppose that the sample $D$ is gathered by some
procedure in which the probability that a star at $(\vx\obs,\vv\obs)$
is included is given by a selection function of the form
$S(\vx\obs,\vv\obs)$.  We assume that we have perfect knowledge of this
$S$; the sometimes-difficult issue of how to construct it in practice
is beyond the scope of the present paper \citep[see,
e.g.,][]{Bovy+12}.
With this $S$, the contribution of the $n^{\rm th}$
star to the likelihood changes from~\eqref{eq:starbystarlik} to
\begin{equation}
  \begin{split}
&\pr(\vx\obs_n,\vv\obs_n,\vZ_n|\vpi,\{\vJ,\vLambda\},\Phi,S)\\
&\quad=\frac{S(\vx\obs_n,\vv\obs_n)\pr(\vx\obs_n,\vv\obs_n,\vZ_n|\vpi,\{\vJ,\vLambda\},\Phi)}
{\sum_{\{z_{km}\}}\int\d\vx\obs\d\vv\obs\,S(\vx\obs,\vv\obs)\pr(\vx\obs,\vv\obs,\{z_{km}\}|\vpi,\{\vJ,\vLambda\},\Phi)},
  \end{split}
\label{eq:Slike1star}
\end{equation}
in which the denominator ensures that the projected likelihood is
correctly normalised.  The treatment of selection functions
$S(\vx^\star,\vv^\star)$ is similar.

\subsection{Distinct stellar populations}

For simplicity we have assumed that the stars in the galaxy are drawn
from a single population.  If the sample~$D$ splits cleanly into, say,
two chemically distinct subpopulations, $D_1$ and $D_2$, then each
population has its own independent DF and the marginal likelihood for
the full sample is simply $\pr(D|\Phi)=\pr(D_1|\Phi)\pr(D_2|\Phi)$.
The more general case in which the properties (e.g., age, metallicity,
$\alpha$-element abundance) of individual stars are ambiguous can be dealt
with by extending the latent variable $z_{nkm}$ introduced in
equation~\eqref{eq:lik2} to indicate the parent DF of each star.

\section{Summary and conclusions}
\label{sec:summary}

The motivation for this paper was to find a Bayesian alternative to
the virial theorem: given a discrete realisation of a galaxy's unknown
DF, what is the unknown potential in which the stars are moving?  The
result, which is encapsulated in equations \eqref{eq:PDgivenPhi}
and~\eqref{eq:sumoverblobs}, follows from marginalising over all
possible equilibrium DFs, adopting a Dirichlet process mixture model
for the prior probability distribution on the DF.  The paper is
essentially a proof-of-concept demonstration that it is feasible to
calculate this marginal likelihood for the idealised case of a
perfect, error-free snapshot of the positions and velocities
$(\vx^\star_n,\vv^\star_n)$ of an unbiased sample of stars from the
galaxy.

The fundamental assumption of the method is that galaxies are in
a steady state.  We know that this is not strictly true, but we can
reasonably expect most galaxies to be sufficiently close to
equilibrium that the steady-state assumption is a good starting point
on which to base more sophisticated time-dependent models
\cite[e.g.,][]{Binney05}.  The next assumption is that galaxy
potentials are integrable and we can map at will between $(\vx,\vv)$
and action-angle coordinates $(\vJ,\vtheta)$ given a potential
$\Phi(\vx)$.  Fortunately, the machinery for constructing such
mappings is already at hand \citep{McMillanBinney08,Binney12,Sanders12}.

We have implicitly assumed that constructing this
$(\vx,\vv)\leftrightarrow(\vJ,\vtheta)$ mapping is expensive.  In
principle, one could attempt to constrain both $\Phi$ and $f$
simultaneously, using, e.g., standard Markov chain Monte Carlo methods
to explore the posterior distribution $\pr(f,\Phi|D)$ of both $\Phi$
and $f$.  Then the constraints on $\Phi$ would come from using the
Markov chain samples to marginalise $f$ (see, e.g., \cite{Bovy+10}
who did this for a restricted family of DFs for the solar-system problem).
Doing this would require that the $(\vx,\vv)\to(\vJ,\vtheta)$ mapping
be constructed anew every time a new trial $\Phi$ is proposed.
Therefore it seems more practical to carry out the marginalisation
over $f$ for each of a range of fixed trial~$\Phi$, leading to
equation~\eqref{eq:PDgivenPhi} for $\pr(D|\Phi)$.

The next step is to apply the ideas presented here to a more realistic
problem.
Perhaps the most promising immediate application would be to
radial velocity surveys of stars in dwarf spheroidal galaxies
\citep[e.g.,][]{Breddels+13} or of globular clusters around massive
galaxies \citep[e.g.,][]{WuTremaine06}; in such systems the selection
effects are relatively straightforward and the method presented in
this paper has the advantage over most others of making no assumptions
about the form of the poorly constrained number-density profile of the
kinematical tracers.  
In \S\ref{sec:gener} we wrote down an expression for the marginalised
likelihood $\pr(D|\Phi)$ when the sample $D$ suffers from real
observational errors and selection biases.  In
Appendix~\ref{sec:exact} we show how this can be reduced to a
sum~\eqref{eq:marglikegen} of
contributions~\eqref{eq:marglikepartition} from different partitions
of the set of stars.  This suggests that an approximation scheme that
uses clustering algorithms to identify the dominant terms in the sum
might be effective.  In developing any such scheme it worth
remembering that we do not really care about the numerical value of
$\pr(D|\Phi)$ itself; we are more interested in the {\it changes} in
$\pr(D|\Phi)$ as we change the assumed potential.

\section*{Acknowledgments}
I thank Scott Tremaine for suggesting the problem of finding a
Bayesian alternative to the virial theorem, and Jo Bovy, Iain Murray,
Ewan Cameron and the anonymous referees for helpful comments.  My
colleagues in the Oxford dynamics group have provided invaluable
constructive criticism throughout all stages of the development of the
work presented here.

\def\aap{A\&A}\def\aj{AJ}\def\apj{ApJ}\def\mnras{MNRAS}\def\araa{ARA\&A}\def\aapr{Astronomy \&
  Astrophysics Review}
\bibliographystyle{mn2e}
\bibliography{sho}

\newpage
\appendix
\onecolumn
\section{The truncated Wishart distribution}
\label{sec:lambdastuff}

This appendix explains how we truncate the improper, uninformative
prior~\eqref{eq:priorlam1} introduced in section~\ref{sec:DPM},
\begin{equation*}
  \pr(\vLambda_k)=B_0|\vLambda_k|^{-\frac12(d+1)},
\tag{\ref{eq:priorlam1}}
\end{equation*}
to eliminate blobs that are ``small'' compared to the cell size
$\Delta J$ and ``large'' compared to the action-space volume $(2J_{\rm
  box})^d$.  The truncated Wishart distribution introduced
here (equ.~\ref{eq:wishart}) reappears
in the calculation of the marginal
likelihood~\eqref{eq:sumoverblobs} 
described in Appendices~\ref{sec:exact} and~\ref{sec:VB} below.

A well-known distribution that is similar to~\eqref{eq:priorlam1} is the
Wishart distribution \citep[e.g.,][]{Press12}, which has density
\begin{equation}
  {\cal W}_0(\vLambda|\vW,\nu)\propto
  |\vLambda|^{\frac12(\nu-d-1)}
  \exp\left[-\textstyle\frac12\tr(\vW^{-1}\vLambda)\right],
\label{eq:wishartorig}
\end{equation}
controlled by the two parameters $\vW$ and $\nu$.
Comparison of \eqref{eq:priorlam1} and~\eqref{eq:wishartorig} suggests
that one way of truncating the uninformative~\eqref{eq:priorlam1} is
by taking $\pr(\vLambda_k)={\cal W}_0(\vW_0,\nu_0)$ with $\nu_0\to0$ and 
$\vW_0=W_0I$, where $I$ is the identity matrix and $W_0$ is related to
the cell size $\Delta J$ through $W_0=(\Delta J)^{-2}$.  This ${\cal
  W}_0(\vW_0,\nu_0)$ is directly 
proportional to the uninformative prior~\eqref{eq:priorlam1} for blobs
that are large compared to $\Delta J$ (i.e., for which
$\tr(\vW_0^{-1}\vLambda_k)\ll1$).  A snag is the Wishart
distribution is normalisable only for $\nu>d-1$.  To remedy this we take
\begin{equation}
  \pr(\vLambda_k)={\cal W}(\vLambda_k|\vW_0,\nu_0),
\label{eq:priorlam}
\end{equation}
where we define the {\it truncated} Wishart distribution~${\cal
  W}(\vLambda|\vW,\nu)$ to be that obtained
from~\eqref{eq:wishartorig} by excluding blobs
with
``volumes'' $|\vLambda|^{-1/2}$ larger
than $(2J_{\rm max})^d$, where $J_{\rm max}\propto J_{\rm box}$:
\begin{equation}
  \label{eq:wishart}
  {\cal W}(\vLambda|\vW,\nu)=
  \begin{cases}
    B(\vW,\nu)|\vLambda|^{\frac12(\nu-d-1)}
    \exp\left[-\textstyle\frac12\tr(\vW^{-1}\vLambda)\right],
    & \hbox{if $|\vLambda|^{-1/2}\lesssim (2J_{\rm max})^d$.}\cr
    0, & \hbox{otherwise}.
  \end{cases}
\end{equation}
The rest of this Appendix is concerned with deriving an explicit
expression for the normalisation constant $B(\vW,\nu)$
that appears in~\eqref{eq:wishart}.
The derivation follows the same lines used to
normalise the conventional, untruncated Wishart distribution ${\cal
  W}_0(\vLambda|\vW,\nu)$ \citep[e.g.,][]{Press12}.

We may assume that the matrix $\vW^{-1}$ is symmetric, so let us
begin by choosing a basis in which
$\vW^{-1}=\diag(w_1^{-1},...,w_d^{-1})$.
$B(\vW,\nu)$ is given by
\begin{equation}
\frac1{B(\vW,\nu)}=
  \int |\vLambda|^{\frac12(\nu-d-1)}
  \exp[-\textstyle\frac12\tr(\vW^{-1}\vLambda)]\,\d\vLambda,
\label{eq:1overB0}
\end{equation}
where the integral is over all positive-definite symmetric matrices
$\vLambda$ that satisfy the constraint $|\vLambda|^{-1/2}\lesssim(2J_{\rm
  max})^d$.  We can express this as an
explicit $\frac12d(d+1)$-dimensional integral by carrying out a Cholesky
decomposition on~$\vLambda$, writing it as 
\begin{equation}
  \vLambda = \vT\vT^T,
\label{eq:cholesky}
\end{equation}
where $\vT$ is a lower triangular matrix whose diagonal elements are
strictly positive, $T_{ii}>0$.  Then the determinant,
\begin{equation}
  |\vLambda|=|\vT\vT^T|=|\vT|^2=\prod_{i=1}^dT_{ii}^2,
\end{equation}
is just the product of these diagonal elements.
We impose the constraint that $|\vLambda|^{-1/2}\lesssim(2J_{\rm
  max})^d$ by considering only $T_{ii}>T_{\rm min}$, where $T_{\rm min}=(2J_{\rm
  max})^{-1}$.  
From~\eqref{eq:cholesky} it is not difficult to show that
\begin{equation}
  \tr(\vW^{-1}\vLambda)=\sum_{i,j}w_i^{-1}T_{ij}^2
\end{equation}
and that
\begin{equation}
  \d\vLambda=2^d\prod_{i=1}^d T_{ii}^{d-i+1}\prod_{j=1}^i\d T_{ij}.
\end{equation}
Then~\eqref{eq:1overB0} becomes
\begin{equation}
  \begin{split}
\frac1{B(\vW,\nu)}&=
2^d\left[\prod_{i=1}^d\int_{T_{\rm min}}^\infty T_{ii}^{\nu-i}
\exp\left(-\frac12w_i^{-1}T_{ii}^2\right)
\d T_{ii}\right]
\prod_{i=1}^d\prod_{j=1}^{i-1}
\int_{-\infty}^\infty\exp\left(-\frac12w_i^{-1}T_{ij}^2\right)
\d T_{ij}\\
&= 2^{\frac12d\nu}|\vW|^{\frac12\nu}\pi^{d(d-1)/4}
\prod_{i=1}^d
\Gamma\left(\frac12(\nu-i+1),\frac{T_{\rm min}^2}{2w_i}\right),
  \end{split}
\label{eq:1overB}
\end{equation}
in which the integral over each of the diagonal elements $T_{ii}$
introduces a lower incomplete Gamma function,
$\Gamma\left(\frac12(\nu-i+1),T_{\rm min}^2/2w_i\right)$.  If
$\nu>d-1$ and the scale set by the parameter $\vW$ is small compared
to $(2J_{\rm max})^d$, then $T_{\rm min}^2/2w_i\to0$ and this reduces
to the usual expression for the normalising constant of the
untruncated Wishart distribution~\eqref{eq:wishartorig}.  We set
$J_{\rm max}$ equal to a few times $J_{\rm box}$.

\section{An exact calculation of the marginal likelihood}

\label{sec:exact}

In this appendix we first derive an expression for the marginal
likelihood $\pr(D|\Phi,\assumptions)$ for the general
case~\eqref{eq:PDgivenPhi2} in which the $N$ stars in the sample~$D$
suffer from observational errors and selection effects.  Then we use
this result to obtain an explicit expression for
$\pr(D|\Phi,\assumptions)$ in the special case of an unbiased,
error-free sample.

Our starting point is the derivation of the Dirichlet process used
in~\S\ref{sec:DP}.  We set up a very fine grid in action space with
$K\to\infty$ cells and use $\vJ_k$ to refer to the location of the
$k^{\rm th}$ cell: a blob is attached to each cell, although most
cells will have zero mass, $\pi_k=0$.  Introducing the latent variable
$\vZ=\{z_{nk}\}$ that indicates whether star~$n$ belongs to cell~$k$,
the marginal likelihood~\eqref{eq:PDgivenPhi2} can be written as
\begin{equation}
  \begin{split}
    \pr(D|\Phi,\assumptions)=\sum_{\vZ}\pr(DZ|\Phi,\assumptions)
  &=
\sum_\vZ\int\d\vpi\pr(\vZ|\vpi)\pr(\vpi)
\prod_{k=1}^K\int\d\vLambda_k\,\pr(\vLambda_k)
\prod_{n=1}^N\left[\pr(\vx^\star_n,\vv^\star_n|\vJ_k,\vLambda_k,\Phi\})\right]^{z_{nk}},
  \end{split}
\label{eq:pinlike}
\end{equation}
where $\pr(\vLambda_k)={\cal W}(\vLambda_k|\vW_0,\nu_0)$ is the
truncated Wishart distribution~\eqref{eq:wishart}.  Substituting for
$\pr(\vZ|\vpi)$ and $\pr(\vpi)$ from \eqref{eq:przpi}
and~\eqref{eq:priorpi} and marginalising $\vpi$ using standard
properties of the Dirichlet distribution, we have that
\begin{equation}
  \begin{split}
  \pr(D|\Phi,\assumptions)
  &=
  \frac{\Gamma(\alpha)}{\Gamma(\alpha+N)}
  \sum_{\vZ} \prod_{k=1}^K
  \frac{\Gamma\left(\frac\alpha K+\bar N_k\right)}
  {\Gamma\left(\frac\alpha K\right)}
  \prod_{k=1}^K\int\d\vLambda_k\,\pr(\vLambda_k)
  \prod_{n=1}^N\left[\pr(\vx^\star_n,\vv^\star_n|\vJ_k,\vLambda_k,\Phi)\right]^{z_{nk}},
  \end{split}
\label{eq:pinlike2}
\end{equation}
where
\begin{equation}
  \label{eq:Nbar}
  \bar N_k\equiv \sum_{n=1}^Nz_{nk}
\end{equation}
is the number of stars that ``belong'' to the $k^{\rm th}$ cell.
Notice that each of the integrals over $\vLambda_k$ is 
1 unless $\bar N_k\ne0$.  So, let us focus our attention on
cells that have $\bar N_k>0$ and rewrite the sum over $\vZ$ as a sum over
partitions\footnote{Recall that a partition of a set $A$ is a division
  into non-overlapping, non-empty subsets of $A$.} of the set
$\{(\vx^\star_1,\vv^\star_1),...,(\vx^\star_N,\vv^\star_N)\}$ into clusters of stars that belong
to the same parent blob $(\vJ_k,\vLambda_k)$, then sum over
the possible locations $\vJ_k$ of the clusters within
each partition.
  That is, having ``pinned'' the cells' locations
$\vJ_k$ to
obtain~\eqref{eq:pinlike} we now unpin them and write 
\begin{equation}
  \begin{split}
    \sum_{\vZ}\prod_{k=1}^K
    &= \sum_{P}
    \sum_{\vJ_1}\cdots\sum_{\vJ_{n_P}},
  \end{split}
  \end{equation}
  where, given a partition $P$ of the $N$ stars
  $\{(\vx^\star_1,\vv^\star_1),...,(\vx^\star_N,\vv^\star_N)\}$, $n_P$
  is the number of elements (i.e., ``clusters'') in $P$ and $\vJ_k$ is
  the location of the $k^{\rm th}$ of the $n_P$ clusters.  In writing
  this sum it is understood that the $\vJ_k$ are distinct.
In the limit $K\to\infty$ each
  $\sum_{\vJ_k}$ becomes $\int\d^d\vJ_k/(\Delta J)^d$, where $\Delta
  J=2J_{\rm box}K^{1/d}$ is the cell size.
  Then, substituting for $\pr(\vJ_k)$
  from~\eqref{eq:priorj} and using $\Gamma(\alpha/K)\to K/\alpha$, our
  general expression for the marginal likelihood is the sum over
  partitions
\begin{equation}
  \begin{split}
  \pr(D|\Phi,\assumptions)
  &=
  \frac{\Gamma(\alpha)}{\Gamma(\alpha+N)}
  \sum_{P} \alpha^{n_p}\prod_{k=1}^{n_P}
  \Gamma\left(\bar N_k\right)
  \pr(D|P_k,\Phi),
  \end{split}
\label{eq:marglikegen}
\end{equation}
where
\begin{equation}
\pr(D|P_k,\Phi)\equiv
  \int\d\vJ_k\d\vLambda_k\pr(\vJ_k)\pr(\vLambda_k)
  \prod_{n=1}^N\left[\pr(\vx^\star_n,\vv^\star_n|\vJ_k,\vLambda_k,\Phi)\right]^{z_{nk}}
\label{eq:marglikepartition}
\end{equation}
is the marginal likelihood of the $\bar N_k$ stars that belong to the
$k^{\rm th}$ cluster of partition~$P$.

For the special situation in which the $(\vx^\star_n,\vv^\star_n)$
constitute an error-free, unbiased snapshot of the stars in the
galaxy, we immediately have that
$\pr(\vx^\star_n,\vv^\star_n|\vJ_k,\vLambda_k,\Phi)=\Blob(\vJ^\star_n|\vJ_k,\vLambda_k)$,
where $\vJ^\star_n$ are the actions of the orbit that passes through
the point $(\vx^\star_n,\vv^\star_n)$ in the assumed potential~$\Phi$.
Then the per-cluster contribution~\eqref{eq:marglikepartition} to the
marginal likelihood becomes
\begin{equation}
  \begin{split}
\pr(D|P_k,\Phi)&\equiv
\sum_{\{z_{nkm}\}}
  \int\d\vJ_k\d\vLambda_k\pr(\vJ_k)\pr(\vLambda_k)
  \prod_{n=1}^N\prod_{m=1}^M\left[{\cal
      N}(\vJ^\star|\vJ_k,\vLambda_k)\right]^{z_{nkm}},
  \label{eq:marglikepartitionexact}
  \end{split}
\end{equation}
in which the sum is over all $M^{\bar N_k}$ assignments $z_{nkm}$ of
the $\bar N_k$ stars to the $M$ Gaussians that make up the blob.
Writing out explicit expressions for the Normal~\eqref{eq:normal} and
truncated Wishart distributions~\eqref{eq:wishart} that appear in this
expression, the integrand is 
\begin{equation}
  \begin{split}
    &\pr(\vJ_k){\cal W}(\vLambda_k|\vW_0,\nu_0)\prod_{n=1}^N\prod_{m=1}^M
      \left[{\cal N}(\vJ^\star_n|R_m\vJ_k,\vLambda_k^{-1})\right]^{z_{nkm}}\\
      &\quad=\pr(\vJ_k)
      \frac{B(\vW_0,\nu_0)}
      {(2\pi)^{\frac12d\bar N_k}}
      {|\vLambda_k|^{\frac12(\bar N_k+\nu_0-d-1)}}
      \exp\left[
    -\frac12\tr(\vW_0^{-1}\vLambda_k)
    -\frac12\sum_{n=1}^N\sum_{m=1}^M
    z_{nkm}(\vJ_n^\star-R_m\vJ_k)^T\vLambda_k(\vJ_n^\star-R_m\vJ_k)
    \right],\\
  \end{split}
\end{equation}
where $B(\vW_0,\nu_0)$ is the normalisation constant~\eqref{eq:1overB}
of the truncated Wishart prior for~$\vLambda_k$.  
Now use the identities
$\vx^TA\vx=\tr(A\vx\vx^T)$ and
$(\vJ_n^\star-R_m\vJ_k)^T\vLambda_k(\vJ_n^\star-R_m\vJ_k) =
(R_m\vJ_n^\star-\vJ_k)^T\vLambda_k(R_m\vJ_n^\star-\vJ_k)$ to complete
the square in the argument of the exponential:
\begin{equation}
  \begin{split}
    \sum_{n=1}^N\sum_{m=1}^Mz_{nkm}
    (\vJ_n^\star-R_m\vJ_k)^T\vLambda_k(\vJ_n^\star-R_m\vJ_k)
  &=    \sum_{n=1}^N\sum_{m=1}^Mz_{nkm}
  \tr\left[\vLambda_k(R_m\vJ_n^\star-\vJ_k)(R_m\vJ_n^\star-\vJ_k)^T\right]\\
  &=    \sum_{n=1}^N\sum_{m=1}^Mz_{nkm}
\tr\left[\vLambda_k((R_m\vJ_n^\star-\bar\vJ_k)+(\bar\vJ_k-\vJ_k))
    ((R_m\vJ_n^\star-\bar\vJ_k)+(\bar\vJ_k-\vJ_k))^T\right]\\
  &=\tr\left[\vLambda_k\bar N_k\bar\vS_k
    +\bar N_k\vLambda_k(\bar\vJ_k-\vJ_k)(\bar\vJ_k-\vJ_k)^T
    \right],
  \end{split}
\label{eq:completesq}
\end{equation}
where in the last two lines we have identified the first few moments
of the actions $\vJ^\star_1,...,\vJ^\star_N$ of the stars that
``belong'' to each of the $n_P$ clusters:
\begin{equation}
  \bar N_{k}=\sum_{n=1}^N\sum_{m=1}^Mz_{nkm},\qquad
  \bar{\vJ}_{k}=\frac1{\bar N_{k}}\sum_{n=1}^N\sum_{m=1}^Mz_{nkm}R_m\vJ^\star_n,\qquad
  \bar \vS_{k}=\frac1{\bar N_{k}}\sum_{n=1}^N\sum_{m=1}^Mz_{nkm}(R_m\vJ^\star_n-\bar\vJ_{k})(R_m\vJ^\star_n-\bar\vJ_{k})^T.
\label{eq:Nkzdefn}
\end{equation}
That is, $\bar N_k$ is the number of stars that belong the $k^{\rm
  th}$ cluster of the partition $P$ (as before), $\bar\vJ_k$ is their
mean action and $\bar\vS_k$ the corresponding covariance matrix.
Introducing $\vW_k^{-1}\equiv\vW_0^{-1}+\bar N_k\bar\vS_k$,
$\nu_k=\nu_0+\bar N_k$ and substituting these results back
into~\eqref{eq:marglikepartitionexact}, the expression for the
contribution of the $k^{\rm th}$ cluster becomes
\begin{equation}
  \begin{split}
  \pr(\vJ^\star_1...\vJ^\star_N|P_k)
  &=
  \frac1{(2J_{\rm box})^d}
  \frac{B(\vW_0,\nu_0)}
  {(2\pi)^{\frac12d\bar N_k}}
  \sum_{\{z_{nkm}\}}
  \int\d\vJ_k\d\vLambda_k
      {|\vLambda_k|^{\frac12(\nu_k-d-1)}}
      \exp\left[-\frac12\tr\left(
          \vLambda_k(\vW_k^{-1}+\bar N_k(\bar\vJ_k-\vJ_k)(\bar\vJ_k-\vJ_k)^T
        \right)\right].
  \end{split}
\end{equation}
Interchanging the order of integration, the integral over $\vJ_k$
is $(2\pi)^{d/2}|\bar 
N_k\vLambda_k|^{-1/2}\exp[-\frac12\tr(\vW_k^{-1}\vLambda_k)]$, leaving
\begin{equation}
  \begin{split}
  \pr(\vJ^\star_1...\vJ^\star_N|P_k)
  &=
  \frac1{(2J_{\rm box})^d}
  \frac{B(\vW_0,\nu_0)}
  {(2\pi)^{\frac12d(\bar N_k-1)}\bar N_k^{d/2}}
  \sum_{\{z_{nkm}\}}
  \int\d\vLambda_k
      {|\vLambda_k|^{\frac12(\nu_k-d-2)}}
      \exp\left[-\frac12\vLambda_k\vW_k^{-1}\right]\\
      &=
  \frac1{(2J_{\rm box})^d}
  \frac{B(\vW_0,\nu_0)}
  {(2\pi)^{\frac12d(\bar N_k-1)}\bar N_k^{d/2}}
  \sum_{\{z_{nkm}\}}
  \frac{1}{B(\vW_k,\nu_k-1)},
  \end{split}
\end{equation}
where we have used equation~\eqref{eq:1overB0} to express the integral
over $\vLambda_k$ as $1/B(\vW_k,\nu_k-1)$.  Substituting this back
into equations \eqref{eq:marglikegen}
and~\eqref{eq:marglikepartition}, our final expression for the
marginal likelihood~\eqref{eq:sumoverblobs} becomes
\begin{equation}
  \pr(\vJ^\star_1...\vJ^\star_N|\assumptions)
  =  \frac{\Gamma(\alpha)}{\Gamma(\alpha+N)}
  \sum_{P}
  \left[\frac{\alpha B(\vW_0,\nu_0)}{(2J_{\rm box})^d}\right]^{n_{P}}
  \prod_{k=1}^{n_{P}}
  \frac{\Gamma\left(\bar N_k\right)}{(2\pi)^{\frac12d(\bar N_k-1)}\bar N_k^{d/2}}
  \left\{\sum_{m_1=1}^M\cdots\sum_{m_N=1}^M
  \frac{1}{B(\vW_k,\nu_k-1)}\right\}.
\label{eq:exactmarglike}
\end{equation}
Identifying $B_0\equiv B(\vW_0,\nu_0)$, notice that the quantity in
square brackets is just the variable $\alpha'$ introduced in
equation~\eqref{eq:alphaprime}.

Given $N$ stars, the number of
partitions~$P$ in the outer sum of~\eqref{eq:exactmarglike} is given
by the {\it Bell number} ${\cal B}_N$, where ${\cal B}_0=1$ and the
${\cal B}_N$ satisfy the recurrence relation \citep[see, e.g.,][]{Rota64}
\begin{equation}
  {\cal B}_{N+1}=\sum_{n=0}^N{N\choose n}{\cal B}_n.
\end{equation}
For $N=10$ stars there are ${\cal B}_{10}=115975$ such partitions to consider,
each of which involves an additional $M^N$ choices of $m$ for each
cluster!  This combinatorial explosion renders this exact calculation
impractical for realistic values of $N$.  Nevertheless, it is feasible to
carry out this sum for $N\le10$ (see \S\S\ref{sec:shodpm} and
\ref{sec:ss}), which provides a
useful check of less expensive, approximate methods.

The terms in the sum~\eqref{eq:exactmarglike} depend on the stars'
actions $\vJ^\star_1,...,\vJ^\star_N$ through the factor
$[B(\vW_k,\nu_k-1)]^{-1}\propto |\vW_k|^{\frac12(\bar
  N_k-1)}\simeq |\bar N_k\bar\vS_k|^{-\frac12(\bar N_k-1)} $,
where $\bar\vS_k$ is the covariance matrix of the stars that belong to
the $k^{\rm th}$ cluster (equ.~\ref{eq:Nkzdefn}).  Therefore the
marginal likelihood~\eqref{eq:PDgivenPhi}
peaks for
choices of potential that produce the sharpest distributions of stars
in action space.

\section{Estimating the marginal likelihood}
\label{sec:VB}

This Appendix explains one way of estimating the value of the (log)
marginal likelihood
\begin{equation}
    P(\vJ^\star)
    \equiv\log\pr(\vJ^\star)=
    \log\left[
      \sum_{\vZ}\int\!\!\int\!\!\int\d\vpi\d\vJ\d\vLambda\pr(\vJ^\star\vZ\vpi\vJ\vLambda)
    \right]
\end{equation}
by using a variational method to find a lower bound.  In this and
subsequent expressions we use $\vJ^\star$ without subscripts to stand
for the full set of the stars' actions
$\{\vJ^\star_1,...,\vJ^\star_N\}$.  Similarly, $\vJ$ and $\vLambda$
without subscripts stand for the blob parameters $\{\vJ_1,...,\vJ_K\}$
and $\{\vLambda_1,...,\vLambda_K\}$, respectively.

Introducing another probability distribution
$Q(\vZ\vpi\vJ\vLambda|\vJ^\star)$, we can
use Jensen's inequality to write
\begin{equation}
  \begin{split}
    P
  &= \log\left[
    \sum_{\vZ}\int\!\!\int\!\!\int \d\vpi\d\vJ\d\vLambda\,
    Q(\vZ\vpi\vJ\vLambda|\vJ^\star)
  \frac{\pr(\vJ^\star\vZ\vpi\vJ\vLambda)}{Q(\vZ\vpi\vJ\vLambda|\vJ^\star)}
  \right]\\
  &\ge \sum_{\vZ}\int\!\!\int\!\!\int\d\vpi\d\vJ\d\vLambda\,
  Q(\vZ\vpi\vJ\vLambda|\vJ^\star)
  \log\frac{\pr(\vJ^\star\vZ\vpi\vJ\vLambda)}{Q(\vZ\vpi\vJ\vLambda|\vJ^\star)}
  \equiv{\cal L}(\vJ^\star).
  \end{split}
\label{eq:lowerbound}
\end{equation}
Using the product rule
$\pr(\vJ^\star\vZ\vpi\vJ\vLambda)=\pr(\vZ\vpi\vJ\vLambda|\vJ^\star)\pr(\vJ^\star)$,
it is easy to see that the difference between the true marginal
likelihood $P$ and the lower bound ${\cal L}$ is given by the
Kullback--Leibler (KL) divergence between $Q$ and $P$,
\begin{equation}
  \hbox{KL}(Q||P)=-\sum_\vZ\int\!\!\int\!\!\int\d\vpi\d\vJ\d\vLambda\,
  Q(\vZ\vpi\vJ\vLambda|\vJ^\star)
  \log\frac{\pr(\vZ\vpi\vJ\vLambda|\vJ^\star)}{Q(\vZ\vpi\vJ\vLambda|\vJ^\star)},
\label{eq:KL}
\end{equation}
which is greater than zero unless
$Q=\pr(\vZ\vpi\vJ\vLambda|\vJ^\star)$.  The idea behind variational
inference is to find a distribution $Q$ that maximises the lower bound
$\cal L$ (thereby minimising $\hbox{KL}(Q||P)$), while simultaneously
leaving the integrals in the expression~\eqref{eq:lowerbound} for
${\cal L}$ tractable.  \cite{MacKay03} explains the origin of this
idea from mean-field theories in statistical physics in which the
partition function is estimated by minimising a variational free
energy.  The treatment below is an adaptation of that presented in
\cite{Bishop06}.

Bearing the need to have a tractable expression for ${\cal L}$, let us
restrict our attention to distributions $Q$ of the factorised form
\begin{equation}
  Q(\vZ\vpi\vJ\vLambda|\vJ^\star)=
  \prod_{n=1}^N\prod_{k=1}^K\prod_{m=1}^MQ_{\vz_{nkm}}(z_{knm}|\vJ^\star)
  Q_{\pi_k}(\pi_k|\vJ^\star)
  Q_{\vJ_k,\vLambda_k}(\vJ_k,\vLambda_k|\vJ^\star).
\label{eq:Qfactorised}
\end{equation}
To keep notation reasonably compact we drop the subscripts and the
explicit dependence on $\vJ^\star$ in these $Q$ factors and use
$Q(\pi_k)$ as shorthand for $Q_{\pi_k}(\pi_k|\vJ^\star)$ and so on.  A
straightforward application of variational calculus shows that the
$Q(\vZ\vpi\vJ\vLambda)$ of the factorised form~\eqref{eq:Qfactorised}
that maximises~${\cal L}$ is given by
  \begin{align}
  \log Q(\vZ)\equiv\sum_{n=1}^N\sum_{k=1}^K\sum_{m=1}^M\log Q(z_{kmn})
  &=\int \d\vpi\int\d\vJ\int\d\vLambda\,
       Q(\vpi)Q(\vJ\vLambda) \log\pr(\vJ^\star\vZ\vpi\vJ\vLambda)
       +\hbox{constant},
\label{eq:optimalqz}
\\
  \log Q(\vpi) \equiv\sum_{k=1}^K\log Q(\pi_k)
  &=\sum_{\vZ} \int\d\vJ\int\d\vLambda\, 
      Q(\vZ)Q(\vJ\vLambda) \log\pr(\vJ^\star\vZ\vpi\vJ\vLambda)
       +\hbox{constant},
\label{eq:optimalqpi}
\\
  \log Q(\vJ\vLambda)\equiv\sum_{k=1}^K\log Q(\vJ_k,\vLambda_k)
  &=\sum_{\vZ} \int\d\vpi\,
      Q(\vZ)Q(\vpi) \log\pr(\vJ^\star\vZ\vpi\vJ\vLambda)
       +\hbox{constant},
\label{eq:optimalqjl}
\end{align}
in which the additive constants are chosen to ensure that each $Q$
factor is correctly normalised and we have made use of the separability
of the particular form of $\log\pr(\vJ^\star\vZ\vpi\vJ\vLambda)$ in
equation~\eqref{eq:probofeverything}.  Notice that the
optimal choices of each of the three factors depends on the other two.
This suggests an iterative scheme in which, starting from an initial
guess for each factor, we cycle through equations \eqref{eq:optimalqz}
to~\eqref{eq:optimalqjl} to update each in turn, repeating until
convergence is reached.  As $\cal L$ is convex with respect to each
$Q$ this scheme is guaranteed to converge.  One can think of it as a
generalisation of the expectation--maximisation algorithm (which in
turn is a generalisation of the Richardson--Lucy algorithm) in which
pointwise estimates of $\pr(\vJ^\star\vZ\vpi\vJ\vLambda)$ are replaced
by estimates of its shape.

\subsection{Expressions for the optimal $Q$ factors}
The integrals that appear on the
right-hand sides of equations \eqref{eq:optimalqz}
  to~\eqref{eq:optimalqjl} are expectations of
  $\log\pr(\vJ^\star\vZ\vpi\vJ\vLambda)$ with respect to different $Q$
  factors.  A convenient shorthand for such expectations is
\begin{align}
  \Exp_{\pi}[f] \equiv \int \d\vpi\, Q(\vpi) f,\\
  \Exp_{\vZ\pi}[f] \equiv \sum_{\vZ}\int \d\vpi\, Q(\vpi)Q(\vZ) f,
\end{align}
and so on, in which the subscripts to $\Exp$ pick out with which $Q$
distributions the expectation of $f$ is to be taken.  Using this notation
equation~\eqref{eq:optimalqz} for $Q(\vZ)$ becomes
\begin{equation}
  \log
  Q(\vZ)=\Exp_{\vpi\vJ\vLambda}[\log\pr(\vJ^\star\vZ\vpi\vJ\vLambda)]+\hbox{constant}.
\end{equation}
Substituting $\pr(\vJ^\star\vZ\vpi\vJ\vLambda)$ from~\eqref{eq:probofeverything} and
absorbing all terms that do not depend on $\vZ$ into the additive
constant gives
\begin{equation}
  \begin{split}
  \log
  Q(\vZ)
  &=\Exp_{\vpi}[\log\pr(\vZ|\vpi)]+\Exp_{\vJ\vLambda}[\log\pr(\vJ^\star|\vZ\vJ\vLambda)]+\hbox{constant}\\
  &=\sum_{n=1}^N\sum_{k=1}^K\sum_{m=1}^Mz_{nkm}\log\rho_{nkm}+\hbox{constant},
  \end{split}
\end{equation}
where
\begin{equation}
  \rho_{nkm}=\Exp_\vpi[\ln\pi_k]+\frac12\Exp_\vLambda[\log|\vLambda_k|]-\frac12d\log(2\pi)
  -\frac12\Exp_{\vJ\vLambda}\left[(\vJ^\star_n-R_m\vJ_{k})^T\vLambda_k(\vJ^\star_n-R_m\vJ_{k})\right].
\label{eq:rhonks}
\end{equation}
Therefore the optimal $Q(\vZ)$ is
\begin{equation}
  Q(\vZ)=\prod_{i=1}^N\prod_{k=1}^K\prod_{m=1}^Mr_{nkm}^{z_{nkm}},
\label{eq:optimalqfinal}
\end{equation}
where the quantities
\begin{equation}
  r_{nkm}=\frac{\rho_{nkm}}{\sum_{k'=1}^K\sum_{m'=1}^M\rho_{nk'm'}},
 \label{eq:rnks}
\end{equation}
known as the ``responsibilities'', are rescaled
versions of $\rho_{nkm}$ chosen to ensure that $Q(\vZ)$ is
correctly normalised.
For later use note that, from equation~\eqref{eq:optimalqfinal},
\begin{equation}
\Exp_{\vZ}[z_{nkm}]=r_{nkm}.
\label{eq:Eznks}
\end{equation}
As in equation~\eqref{eq:Nkzdefn} of 
Appendix~\ref{sec:exact} above we introduce the following expressions for the first few
moments of the data $\vJ^\star_1,...,\vJ^\star_N$ that ``belong'' to
each of the $K$ blobs in the model:
\begin{equation}
  \bar N_{k}=\sum_{n=1}^N\sum_{m=1}^Mr_{nkm},\qquad
  \bar{\vJ}_{k}=\frac1{\bar N_{k}}\sum_{n=1}^N\sum_{m=1}^Mr_{nkm}R_m\vJ^\star_n,\qquad
  \bar \vS_{k}=\frac1{\bar N_{k}}\sum_{n=1}^N\sum_{m=1}^Mr_{nkm}(R_m\vJ^\star_n-\bar\vJ_{k})(R_m\vJ^\star_n-\bar\vJ_{k})^T.
\label{eq:Nkdefn}
\end{equation}
Notice that the $r_{nkm}$ depend on the values of the three
expectations that appear in~\eqref{eq:rhonks}, which in turn depend on
the choice of $Q(\vpi)$ and $Q(\vJ\vLambda)$.  We now turn to finding
the optimal choices for these two distributions.

Applying the same procedure to
equation~\eqref{eq:optimalqpi} for $Q(\vpi)$, we have that
\begin{equation}
  \begin{split}
  \log
  Q(\vpi)&=\Exp_{\vZ\vJ\vLambda}[\log\pr(\vJ^\star\vZ\vpi\vJ\vLambda)]\\
&=\log\pr(\vpi)+\Exp_{\vZ}[\log\pr(\vZ|\vpi)]+\hbox{constant}\\
&=\log\pr(\vpi)+\sum_{n=1}^N\sum_{k=1}^K\sum_{m=1}^M\Exp_\vZ[z_{nkm}]\log\pi_k+\hbox{constant}.
  \end{split}
\end{equation}
Taking $\pr(\vpi)$ from~\eqref{eq:priorpi}, substituting
$\Exp_{\vZ}[z_{nkm}]=r_{nkm}$ from~\eqref{eq:Eznks}
and then identifying
the quantity $\bar N_k$ introduced in~\eqref{eq:Nkdefn} in the resulting sum over
$r_{nkm}$, the optimal $Q(\vpi)$ is clearly a Dirichlet
distribution~(equ.~\ref{eq:dirichlet}),
\begin{equation}
  Q(\vpi)=\Dir(\vpi|\valpha)
\end{equation}
in which $\alpha_k=\alpha_0+\bar N_k$.

Similarly, the optimal choice of the remaining factor $Q(\vJ\vLambda)$ is
\begin{equation}
  \begin{split}
  \log
  Q(\vJ\vLambda)&=\Exp_{\vZ\vpi}[\log\pr(\vJ^\star\vZ\vpi\vJ\vLambda)]+\hbox{constant}\\
  &=\Exp_{\vZ}[\log\pr(\vJ^\star|\vZ\vJ\vLambda)]+\sum_{k=1}^K\log\pr(\vJ_k)
  +\sum_{k=1}^K\log\pr(\vLambda_k)+\hbox{constant}\\
  &=\sum_{k=1}^K\sum_{n=1}^N\sum_{m=1}^M\Exp_{\vZ}[z_{nkm}]\log{\cal
    N}(\vJ^\star_n|R_m\vJ_{k},\vLambda_k)
  +\sum_{k=1}^K\log{\cal W}(\vLambda_k|\vW_0,\nu_0)
+\hbox{constant}.
  \end{split}
\end{equation}
Writing out explicit expressions for the normal and Wishart
distributions that appear here, gathering together terms involving
each $\vJ_k$ and using the identities
$\vx^TA\vx=\tr(A\vx\vx^T)$ and
$(\vJ_n^\star-R_m\vJ_k)^T\vLambda_k(\vJ_n^\star-R_m\vJ_k) =
(R_m\vJ_n^\star-\vJ_k)^T\vLambda_k(R_m\vJ_n^\star-\vJ_k)$ to complete
the square in the argument of the exponential (see also
equ.~\ref{eq:completesq} above)
and simplifying
gives $Q(\vJ,\vLambda)=\prod_{k=1}^KQ(\vJ_k|\vLambda_k)Q(\vLambda_k)$ with
\begin{equation}
\begin{split}
  Q(\vJ_k|\vLambda_k)&=
  \begin{cases}
    \pr(\vJ_k), & \hbox{if $\bar N_k\le d$},\\
    {\cal N}\left(\vJ_k|\bar\vJ_k,(\bar N_k\vLambda_k)^{-1}\right),
    & \hbox{otherwise}.
  \end{cases}\\
  Q(\vLambda_k)&=
  {\cal W}\left(\vLambda_k|\vW_k,\nu_k\right),
\end{split}
\label{eq:QJL}
\end{equation}
in which
\begin{equation}
  \begin{split}
  \vW_k^{-1}&=\vW_0^{-1}+\bar N_k\bar\vS_k,\\
  \nu_k&=\nu_0+\bar N_k,\label{eq:nuk}
  \end{split}
\end{equation}
and $\bar N_k$, $\bar\vJ_k$ and $\bar \vS_k$ are the
responsibility-weighted moments of the data defined
in~\eqref{eq:Nkdefn}.  We note that the expression~\eqref{eq:QJL} for
$Q(\vJ_k|\vLambda_k)$ is, strictly speaking, the optimal choice only
for the cases $\nu_k\ll1$ or $\nu_k>d$, but it suffices for the following.


\subsection{Algorithm for finding the best $Q$}
\label{sec:VBsteps}

Having $Q(\vpi)$ and $Q(\vJ\vLambda)$ we are now in a position to
calculate all of the expectations that appear in the
expression~\eqref{eq:rhonks} for $\rho_{nkm}$ that determines
$Q(\vZ)$.  Applying standard properties of the Normal, Wishart and
Dirichlet distributions, the relevant results are
\begin{align}
  \Exp_{\vJ_k\vLambda_k}\left[(\vJ^\star_n-R_m\vJ_k)^T\vLambda_k(\vJ^\star_n-R_m\vJ_k)\right]
  &=d\bar N_k^{-1}+\nu_k(R_m\vJ^\star_n-\bar\vJ_k)^T\vW_k(R_m\vJ^\star_n-\bar\vJ_k),\label{eq:expjl}\\
  \log\tilde\Lambda_k\equiv
  \Exp_\vLambda[\log|\vLambda_k|]&=\sum_{i=1}^d\psi\left(\frac12(\nu_k+1-i)\right)
  +d\log2+\log|\vW_k|,\label{eq:Lambdat}\\
  \log\tilde\pi_k\equiv \Exp_{\vpi}[\log\pi_k]&=\psi(\alpha_k)-\psi\left(\sum_k\alpha_k\right),\label{eq:pit}
\end{align}
where the digamma function $\psi(z)\equiv \d\log\Gamma(z)/\d z$ and we
have assumed that $\bar N_k>d$.
Substituting into~\eqref{eq:rhonks} gives, finally,
\begin{equation}
  \rho_{nkm}=\tilde\pi_k\tilde\Lambda_k^{1/2}
  \exp\left[
    -\frac d{2\bar N_k}
    -\frac12\nu_k(R_m\vJ^\star_n-\bar\vJ_k)^T\vW_k(R_m\vJ^\star_n-\bar\vJ_k)
    \right].
\label{eq:rhonks2}
\end{equation}

An algorithm for finding the optimal $Q(\vZ\vpi\vJ\vLambda)$ is to
alternately (i) update $Q(\vZ)$ given $Q(\vpi)$ and
$Q(\vJ\vLambda)$, then (ii) update $Q(\vpi)$ and $Q(\vJ\vLambda)$ given
this new $Q(\vZ)$.  More explicitly, these two alternating steps are:
\begin{enumerate}
\item Having
  estimates of $\alpha_k$, $\bar N_k$, $\bar\vJ_k$, $\vW_k$ and $\nu_k$,
  use equations~\eqref{eq:rnks} and \eqref{eq:Lambdat} to~\eqref{eq:rhonks2} to calculate
  the responsibilities $r_{nkm}$.
\item Plug these $r_{nkm}$ into equation~\eqref{eq:Nkdefn} to obtain
  updated values for the
  responsibility-weighted moments $\bar N_k$, $\bar\vJ_k$ and $\bar \vS_k$.  Set
  $\alpha_k=\alpha_0+\bar N_k$.  Use equations \eqref{eq:nuk}
  to update  $\vW_k$ and $\nu_k$.
\end{enumerate}
We use the $K$-means algorithm \citep{Bishop06} to initialise this
procedure.  The simplest way of checking for convergence is by
examining the rate of increase of the lower bound~${\cal L}$.

\subsection{Evaluation of the lower bound ${\cal L}$}

From equation~\eqref{eq:lowerbound} the lower bound
\begin{equation}
  \begin{split}
  {\cal L}&=\sum_{\vZ}\int\d\vpi\int\d\vJ\int\d\vLambda\,
  Q(\vZ\vpi\vJ\vLambda)
  \log\left\{\frac{\pr(\vJ^\star\vZ\vpi\vJ\vLambda)}{Q(\vZ\vpi\vJ\vLambda)}\right\}\\
  &=\Exp[\log\pr(\vJ^\star\vZ\vpi\vJ\vLambda)]-\Exp[\log Q(\vZ\vpi\vJ\vLambda)]\\
  &=\Exp_{\vZ\vJ\vLambda}[\log\pr(\vJ^\star|\vZ\vJ\vLambda)]
  +\Exp_{\vZ\vpi}[\log\pr(\vZ|\vpi)]
  +E_{\vpi}[\log\pr(\vpi)]
  +\Exp_{\vJ\vLambda}[\log\pr(\vJ)]
  +\Exp_{\vJ\vLambda}[\log\pr(\vLambda)]\\
  &\quad-\Exp_{\vZ}[\log Q(\vZ)]-\Exp_{\vpi}[\log Q(\vpi)]-\Exp_{\vJ\vLambda}[\log Q(\vJ\vLambda)].
  \end{split}
  \label{eq:margliksumexp}
\end{equation}
The expectations~\eqref{eq:margliksumexp} are easy to work out with
the aid of the relations proved above.  For example, taking $\pr(\vJ^\star|\vZ\vJ\vLambda)$
from~\eqref{eq:likjstar} together with the expectations already
worked out in~\eqref{eq:rhonks} and~\eqref{eq:expjl} gives
\begin{equation}
  \Exp_{\vZ\vJ\vLambda}[\log\pr(\vJ^\star|\vZ\vJ\vLambda)]
  =\frac12\sum_{k=1\atop\bar N_k>0}^K\bar N_k\left\{
    \log\tilde\Lambda_k-\nu_k\tr(\vS_k\vW_k)
    -d\log(2\pi)\right\}.
\end{equation}
Similarly, taking $\pr(\vZ|\vpi)$ from~\eqref{eq:przpi} and
$\pr(\vpi)$ from~\eqref{eq:priorpi} together with~\eqref{eq:pit} for
$\Exp_{\vpi}[\log\pi_k]$ gives
\begin{align}
  \Exp_{\vZ\vpi}[\log\pr(\vZ|\vpi)]&=\sum_{k=1}^K\bar N_k\log\tilde\pi_k,
  \\
  \Exp_{\vpi}[\log\pr(\vpi)]&=\log
  C(\valpha_0)+(\alpha_0-1)\sum_{k=1}^K\log\tilde\pi_k.
\end{align}
The other contributions to $\cal L$ are
\begin{align}
\Exp_{\vJ\vLambda}[\log \pr(\vJ)]&=-Kd\log(2J_{\rm box}),\\
  \Exp_{\vJ\vLambda}[\log\pr(\vLambda)]
  &=\sum_{k=1}^K\left\{
    \log B(\vW_0,\nu_0)+\frac{\nu_0-d-1}2\log\tilde\Lambda_k
  -\frac12\nu_k\tr(\vW_0^{-1}\vW_k)\right\},\\
  \Exp_\vZ[\log Q(\vZ)]&=\sum_{n=1}^N\sum_{k=1}^K\sum_{m=1}^Mr_{nkm}\log r_{nkm},
  \\
  \Exp_\vpi[\log  Q(\vpi)]
  &=\log C(\valpha)+\sum_{k=1}^K(\alpha_k-1)\log\tilde\pi_k,
  \\
  \Exp_{\vJ\vLambda}[\log Q(\vJ\vLambda)]
  &=\sum_{k=1}^K  \Exp_{{\vJ_k}{\vLambda_k}}[\log (Q(\vJ_k|\vLambda_k)Q(\vLambda_k)],\\
  \Exp_{{\vJ_k}{\vLambda_k}}[\log Q(\vJ_k|\vLambda_k)Q(\vLambda_k)]
  &=-H[Q(\vLambda_k)]+
  \begin{cases}
    -d\log(2J_{\rm box}), & \hbox{if $\bar N_k\le d$},\\
    \frac12(\log\tilde\Lambda_k+d\log\bar N_k)
    -\frac12d(1+\log2\pi), & \hbox{otherwise}, \\
  \end{cases}
  \\
  H[Q(\vLambda_k)]&=-\log B(\vW_k,\nu_k)
  -\frac{\nu_k-d-1}2\log\tilde\Lambda_k
  +\frac12\nu_kd.
\end{align}
Most of these terms cancel, leaving
\begin{equation}
  {\cal L}=
  \log\left(\frac{C(\valpha_0)}{C(\valpha)}\right)
  +
  \sum_{k=1\atop\bar N_k>d}^K{\cal L}_k
  -\sum_{n=1}^N\sum_{k=1}^K\sum_{m=1}^Mr_{nkm}\log r_{nkm},
\label{eq:simplik}
\end{equation}
in which each blob with $\bar N_k>d$ contributes a term
\begin{equation}
  {\cal L}_k=\log\left(\frac{B(\vW_0,\nu_0)}{B(\vW_k,\nu_k)}\right)
  -\frac12
    \log\tilde\Lambda_k
    +\frac12d\left[1-\log\bar N_k
    -2\log(2J_{\rm box})\right].
\end{equation}
Although it is not immediately obvious, this expression for $\cal L$
is very similar to {\it one} of the terms that appear in the sums
over partitions $P'$ in the exact expression~\eqref{eq:exactmarglike}
for the marginal likelihood.  To show this, take $C$ from
equ~\eqref{eq:Cdefn} and use the approximation that
$\Gamma(\alpha/K)\to (\alpha/K)^{-1}$ when $K$ is large.  Then the
first term in~${\cal L}$ becomes
  \begin{equation}
    \frac{C(\valpha_0)}{C(\valpha)}
    =\frac{\Gamma(\alpha)}{\Gamma(\alpha+N)}
  \prod_{k=1\atop\bar N_k>d}^K
  \frac{\Gamma(\frac\alpha K+\bar N_k)}{\Gamma(\frac\alpha K)} 
  \to\frac{\Gamma(\alpha)}{\Gamma(\alpha+N)}
  \prod_{k=1\atop\bar N_k>d}^K\frac\alpha K{\Gamma(\bar N_k)}
  =\frac{\Gamma(\alpha)}{\Gamma(\alpha+N)}\left(\frac\alpha K\right)^{K_+}
  \prod_{k=1\atop\bar N_k>d}^K\Gamma(\bar N_k),
  \label{eq:Crat}
  \end{equation}
as $K\to\infty$, where $K_+$ is the number of occupied blobs
with $\bar N_k>d$. 
The $Q$ factors in the variational Bayes
algorithm tend to converge on a local maximum of the
distribution.  The maximum is degenerate, however, as can be seen by
permuting the $k$ indices of each blob: in general we will have
$K_+$ blobs with distinct $(\vJ_k,\vLambda_k)$ plus $K-K_+$ identical
blobs with zero mass.  As an approximate way
of accounting for these ``missing'' permutations in the
integral~\eqref{eq:lowerbound}, we simply add a term
$\log(K!/(K-K_+)!)$ to ${\cal L}$.  This 
cancels out the stray $K^{-K_+}$ factor in
equation~\eqref{eq:Crat}.
With this correction, the approximate lower bound on the marginal
likelihood becomes 
\begin{equation}
  \begin{split}
\exp\left[{\cal L}+K_+\log K\right] &=
\exp\left[-\sum_{nkm}r_{nkm}\log r_{nkm}\right]
  \frac{\Gamma(\alpha)}{\Gamma(\alpha+N)}
  \left(\frac{\alpha B(\vW_0,\nu_0)}{K(2J_{\rm box})^d}\right)^{K_+}
  \prod_{k=1\atop\bar N_k>d}^K
  \frac{\Gamma(\bar N_k)}{\bar N_k^{\frac12d}}
  \frac1{\tilde\Lambda_k^{1/2}B(\vW_k,\nu_k)}\\
  &=
\exp\left[-\sum_{nkm}r_{nkm}\log r_{nkm}\right]
  \frac{\Gamma(\alpha)}{\Gamma(\alpha+N)}
  \left(\frac{\alpha B(\vW_0,\nu_0)}{(2J_{\rm box})^d}\right)^{K_+}
  \prod_{k=1\atop\bar N_k>d}^K
  \frac{\Gamma(\bar N_k)}{\bar N_k^{\frac12d}}
  \frac{\hbox{[$\bar N_k$-dependent factors]}}{B(\vW_k,\nu_k-1)},
  \end{split}
\label{eq:approxexact}
\end{equation}
in which we have used~\eqref{eq:1overB} and~\eqref{eq:Lambdat} to write
$\tilde\Lambda_k^{1/2}B(\vW_k,\nu_k)$ as $B(\vW_k,\nu_k-1)$ times some
$\bar N_k$-dependent factors.  Apart from these factors and a
related contribution from the entropic $r_{nkm}\log
r_{nkm}$ prefactor, the result is identical to the contribution made
to the exact result~\eqref{eq:exactmarglike} by a single partition
$P'$ with a specific choice of reflections $(m_1,...,m_{n_{P'}})$.

This shows that the variational estimate is good provided: (a) the
stars divide cleanly into distinct clusters in action space so that the
exact marginal likelihood~\eqref{eq:exactmarglike} is dominated by a
single partition~$P'$; and (b) the two-step algorithm given
in Section~\ref{sec:VBsteps} successfully finds this $P'$.
The smaller the value of the concentration parameter $\alpha'=\alpha
B(\vW_0,\nu_0)/(2J_{\rm max})^d$, the more likely this condition is to
be satisfied.
For the purposes of the present paper, however, we do not strictly
need the estimate to be ``good'' in this sense; it is more important
that the estimate accurately captures {\it changes} in the marginal
likelihood as changes in the trial potential modify the stars' actions
$\{\vJ_1^\star(\vx_1^\star,\vv_1^\star|\Phi),...,\vJ_N^\star(\vx_N^\star,\vv_N^\star|\Phi)\}$.
Perhaps the most obvious example of a situation in which the
estimate~\eqref{eq:approxexact} fails is one in which changing the
potential changes the number $K_+$ of distinct clusters.

\label{lastpage}
\end{document}